\documentstyle[amsfonts,12pt]{article}
\def\theequation{\thesection.\arabic{equation}}
\title{}\date{}

\title{Non-Abelian Vortices in Supersymmetric Gauge Field Theory via Direct
Methods}

\author{Elliott H. Lieb\footnote{Work partially supported by U. S. National
Science Foundation grant PHY -- 0965859.}
\\Department of Mathematics\\Princeton University\\Princeton, New Jersey
08540, USA\\ \\
Yisong Yang \\
Department of Mathematics\\Polytechnic Institute of New York 
University\\Brooklyn, New York 11201, USA}
\newcommand{\bfR}{{\Bbb R}}

\newcommand{\uu}{\underline}
\newcommand{\bfZ}{{\Bbb Z}}

\def\Int{\mathop{\rm Int}}
\def\Re{\mathop{\rm Re}}
\def\Im{\mathop{\rm Im}}

\newcommand{\n}{\noindent}
\newcommand{\p}{\hspace*{\parindent}}

\newtheorem{oldtheorem}{Theorem}[section]
\newtheorem{oldassertion}[oldtheorem]{Assertion}
\newtheorem{oldproposition}[oldtheorem]{Proposition}
\newtheorem{oldremark}[oldtheorem]{Remark}
\newtheorem{oldlemma}[oldtheorem]{Lemma}
\newtheorem{olddefinition}[oldtheorem]{Definition}
\newtheorem{oldclaim}[oldtheorem]{Claim}
\newtheorem{oldcorollary}[oldtheorem]{Corollary}

\newenvironment{theorem}{\begin{oldtheorem}$\!\!\!${\bf.}}{\end{oldtheorem}}

\newbox\qedbox
                        {\hfill\penalty10000\copy\qedbox\par\medskip}

\newcommand{\dd}{\mbox{d}}\newcommand{\D}{\cal O}
\newcommand{\ee}{\end{equation}}
\newcommand{\be}{\begin{equation}}\newcommand{\bea}{\begin{eqnarray}}
\newcommand{\eea}{\end{eqnarray}}
\newcommand{\ii}{\mbox{i}}\newcommand{\e}{\mbox{e}}
\newcommand{\pa}{\partial}\newcommand{\Om}{\Omega}
\newcommand{\vep}{\varepsilon}
\newcommand{\Bd}{B_{\mbox{\small d}}}\newcommand{\Wd}{W_{\mbox{\small
d}}}\newcommand{\Bv}{B_{\mbox{\small v}}}\newcommand{\Wv}{W_{\mbox{\small
v}}}
\newcommand{\nn}{\nonumber}
\newcommand{\F}{\vec{F}}
\newcommand{\Ph}{\vec{\phi}}
\newcommand{\A}{\vec{A}}
\newcommand{\bfph}{{\bf \phi}}
\newcommand{\lm}{\lambda}\newcommand{\sd}{\mbox{\tiny d}}

\begin{document}
\maketitle

\let\thefootnote\relax\footnotetext{\copyright\, 2011 by
the authors. This paper may be reproduced, in its entirety, for
non-commercial purposes.}

\begin{abstract}
Vortices in supersymmetric gauge field theory are important constructs in a
basic conceptual phenomenon commonly referred to as
 the dual Meissner effect which is responsible for color confinement. Based
on a direct minimization approach, we present a series of
sharp existence and uniqueness theorems for the solutions of some
non-Abelian vortex equations governing color-charged multiply distributed
flux tubes, which provide an essential mechanism for
linear confinement. Over a doubly periodic domain, existence results are
obtained under  explicitly stated necessary and sufficient conditions that
relate the size of the domain,
the vortex numbers, and the underlying physical coupling
parameters of the models. Over the full plane, existence results are valid
for arbitrary vortex numbers and coupling parameters. In all cases,
solutions are unique.
\end{abstract}


\section{Introduction}
\setcounter{equation}{0}

A fundamental puzzle in physics, known as the quark confinement, is that
quarks, which make up elementary particles such as mesons and baryons,
cannot be observed in
isolation. A well accepted confinement mechanism, known as the linear
confinement model, states that, when one tries to separate a pair of quarks,
such as a quark and an anti-quark constituting a meson, the
energy consumed would grow linearly with respect to the the separation
distance between the quarks so that it would require an infinite amount of
energy in order to
split the pair. The quark and anti-quark may be regarded as a pair of source
and sink of color-charged force fields. The source and sink interact through
color-charged fluxes
which are screened in the bulk of space but form thin tubes in the form of
color-charged vortex-lines so that the strength of the force remains
constant over arbitrary distance, resulting in
a linear dependence relation for the potential energy with regard to the
separation distance. Such a situation is similar to that of a magnetic
monopole and
anti-monopole pair immersed in a type-II superconductor. The magnetic fluxes
mediating the interacting monopoles are not governed by the Maxwell
equations,
which would otherwise give rise to an inverse-square-power law type of decay
of the forces and lead to non-confinement, but rather by the
Ginzburg--Landau equations, which produce
thin vortex-lines, known as the Abrikosov vortices or the Nielsen--Olesen
strings. The repulsion of the magnetic field in the bulk region
of the superconductor is due to the Meissner effect and
the partial magnetic penetration of the superconductor in the form of
vortex-lines is a consequence of type-II superconductivity. Thus, one may
conclude that
a monopole and anti-monopole immersed in a type-II superconductor would be
linearly confined.

Inspired by the above-described monopole confinement in a type-II
superconductor, Mandelstam \cite{Man1,Man2}, Nambu \cite{Nambu}, and 't
Hooft \cite{tH1,tH2} proposed in the 1970s
that the ground state of quantum-chromodynamics (QCD) is a condensate of
chromomagnetic (color-charged) monopoles, causing the chromoelectric fluxes
between quarks
to be squeezed into narrowly formed tubes or vortex-lines, similar to the
electron condensation
in the bulk of a superconductor, in the form of the Cooper pairs,  resulting
in
the formation of magnetic flux-tubes or vortex-lines which mediate the
interaction between monopoles, following a non-Abelian version of the
Meissner effect,
called the `dual Meissner effect'  \cite{Del,GSY,KSW,Suzuki}, which is
responsible for the screening of chromoelectric fluxes \cite{ShY2,ShY}.

Interestingly, although finite-mass monopoles of the 't Hooft \cite{tH} and
Polyakov \cite{Pol} type in non-Abelian gauge theory have long been
demonstrated to exist \cite{Actor,Go}, the
magnetic fluxes are Coulomb-like which spread out radially and will not give
rise to confinement. In 1994, Seiberg and Witten \cite{SW} came up with an
${\cal N}=2$
supersymmetric gauge-field-theoretical formalism of non-Abelian monopole
condensation and studied its implications to color confinement. Later, in
1997,
Hanany, Strassler, and Zaffaroni \cite{HSZ} showed that the flux tubes or
strings produced in the Seiberg--Witten formalism are of the
Abrikosov--Nielsen--Olesen (Abelian) type
\cite{Ab,GST,NO,T1,T2}
which are not exactly what anticipated in QCD \cite{HSZ,ShY,ShY2}. In 2002,
Marshakov and Yung \cite{MY} constructed Abelian vortices in a softly broken
${\cal N}=2$ supersymmetric QCD (SQCD)
model
and showed that,
although
confinement is due to Abelian flux tubes, the multiplicity of the meson
spectrum is the
same as expected in a theory with non-Abelian confinement. In 2003, Hanany
and Tong \cite{HT} derived a broad class of
non-Abelian vortex equations and computed the dimensions of the associated
moduli spaces, and Auzzi, Bolognesi, Evslin, Konishi, and Yung \cite{Auzzi}
analyzed non-Abelian vortices and confinement in ${\cal N}=2$ SQCD in the
context of non-Abelian superconductors. Since then, the subject of
non-Abelian vortices, monopole condensation, and confinement has been
extensively developed \cite{EF,EFN,EI,GJK,HT2004,ShY2004,ShY-vortex}. See
\cite{Eto-survey,Gr,Kon-survey,Shifman-survey,ShY2,ShY} for surveys and
further literature.
Mathematically, these studies unveil a broad spectrum of systems of elliptic
equations with exponential nonlinearities and rich properties and
structures, which present new challenges.

Recently, C. S. Lin and one of the authors (Y. Y.) carried out a systematic
study \cite{LY09,LY} of the multiple vortex equations obtained in
\cite{EFN,EI,Eto-survey,GJK,ShY2004,ShY-vortex,ShY2,ShY}.
A series of sharp existence and uniqueness theorems were established. The
methods used include monotone iterations, {\em a priori} estimates and
degree-theory argument, and constrained minimization.
In the present paper, we do two things. One is to develop and prove another
series of sharp existence and uniqueness theorems for the multiple vortex
equations derived in \cite{Auzzi,AK}, which are not covered in
\cite{LY09,LY}. The second thing is to develop a methodology that has not
normally been used for these kinds of problems,
over doubly periodic domains, which are often more difficult to approach due
to the appearance of some integral constraints naturally associated with the
equations. This is the
 highly efficient direct minimization approach which enables us to
identify the key analytic ingredients and pursue a complete understanding of
the problems almost immediately. As a by-product, such an approach also
provides a constructive method
for solutions. It is hopeful that our method may be explored further to
study various multiple vortex equations, arising in non-Abelian gauge field
theory, of more difficult structures.
It should be noted that, after solving the problems here by the direct
method, we gained true insight to solve them by the usual constrained
minimization method. But we were not successful in all cases. Thus, the
direct method,
besides being simpler, is sometimes the only constructively workable one.

The content of the rest of the paper is outlined as follows.

In Section 2, we recall the SQCD multiple vortex equations
of Auzzi,  Bolognesi,  Evslin,  Konishi, and Yung \cite{Auzzi} obtained in
2003 in which vortices are induced from three complex scalar fields.
We then state our existence and uniqueness theorem, In the next two
sections, we give the proofs for the various parts of the theorem.
In Section 5, we turn our attention to a study of the multiple vortex
equations derived by Auzzi and Kumar \cite{AK} in a supersymmetric
Chern--Simons--Higgs theory
formulated by Aharony, Bergman, Jafferis and Maldacena  \cite{ABJM}, known
as the ABJM model. In this problem, vortices are generated from $m$ complex
scalar fields and the governing elliptic
system consists of $m$ equations. Again, we are able to obtain a sharp
existence and uniqueness theorem for solutions over a doubly periodic domain
and the full plane. Proofs
of results are sketched in the subsequent two sections, as earlier.

In Section 8, we further illustrate how our direct methods may be used in
tackling other problems of similar
structures. Specifically, in \S 8.1, we revisit the $SO(2N)$ BPS vortex
equations of Gudnason, Jiang, and Konishi \cite{GJK}, studied in \cite{LY}
where solutions are constructed
by a constrained minimization method when the total vortex number $n$ does
not exceed $3$ and existence of solutions is established for arbitrary $n$
using a degree-theory argument.

The main difficulty encountered in \cite{LY} is an extra term in one of the
equations
that makes it hard to resolve the constraints explicitly, which may be flown
away by a homotopy flow. Here is an example where the direct method
seems to be the only (constructively) successful one. With it we are able to
obtain an existence proof
for an arbitrary $n$. In fact, we will carry out our study in the most
general situation where vortices are induced from the sets of zeros of the
two complex scalar fields of
the model.

In \S 8.2, we present a sharp existence and uniqueness theorem for the
multiple vortex equations obtained by Marshakov and Yung \cite{MY} in 2002,
which may be regarded
as the earliest non-Abelian SQCD vortex equations for which the vortex-lines
are taking values in the Cartan subalgebra of $SU(3)$ and, also,  the
starting point of the later development
of the subject of non-Abelian vortices and monopoles in SQCD and their
applications to color confinement. The method is again centered around
direct minimization.

\section{Vortices in Yang--Mills--Higgs theory}
\setcounter{equation}{0}

Following Auzzi,  Bolognesi,  Evslin,  Konishi, and Yung \cite{Auzzi}, the
Yang--Mills--Higgs action hosting the gauge field theory undergoing the
spontaneous symmetry breaking
\be
SU(N)\to SU(N-1)\times U(1),
\ee
within the context of the critical BPS coupling, assumes the form
\bea\label{S}
S&=&
\int\left\{-\frac1{4g^2}F_{\mu\nu}^{a}F^{a\mu\nu}-\frac1{4e^2}F_{\mu\nu}F^{\mu\nu}+(\nabla_\mu
q_A)^\dagger\nabla^\mu q^A\right.\nn\\
&&\quad\left.-\frac{g^2}2({q}^\dagger_A t^a
q^A)^2-\frac{e^2}{4K(K+1)}(q^\dagger_A q^A-K\xi)^2\right\}\,\dd x,
\eea
where $K=N-1$, the index $a=1,2,\cdots,K^2-1$, labels the group generators
$\{t^a\}$  of $SU(K)$, $g,e>0$ are the $SU(K)$ and $U(1)$ gauge-field
coupling constants, respectively,
$\xi>0$ determines the vacuum expectation value of the quark fields $q^A$
lying in the fundamental representation of $SU(K)\times U(1)$,
$A=1,2,\cdots,N_{\mbox{\tiny flavor}}$ runs
over the quark flavors,
\be
\nabla_\mu=\pa_\mu-\ii A_\mu^a t^a-\ii A_\mu t^0,\quad
t^0=\frac1{\sqrt{2K(K+1)}}\left(\begin{array}{cc}{\bf
1}_K&0\\0&-K\end{array}\right),
\ee
denotes the gauge-covariant derivative, with ${\bf1}_m$ the $m\times m$
identity matrix,
and the Minkowski spacetime is of the signature $(+---)$. As a consequence
of the Bogomol'nyi reduction \cite{Bo,JT} for static vortex solutions,
the BPS vortex equations  \cite{Auzzi} are of the form
\bea
\frac1{g^2}F_{12}^a+ (q_A^\dagger t^a q^A)&=&0,\quad a=1,2,\cdots,
K^2-1,\label{BPS1} \\
\frac{1}{e^2}F_{12}+(q_A^\dagger q^A-K\xi)&=&0,\label{BPS2}\\
\nabla_1 q^A+\ii \nabla_2 q^A&=&0,\quad A=1,2,\cdots, N_{\mbox{\tiny
flavor}}.\label{BPS3}
\eea

In the specific situation of $N=4$ so that the unbroken symmetry is given by
the group $SU(3)\times U(1)$, the non-Abelian vortex solutions may be
described by gauge fields solely given in the $a=3,8$ (these are the Cartan
subalgebra indices in the Gell-Mann matrix representation) and the $U(1)$
sectors, and the quark fields are represented by the complex matrix
\be \label{q}
(q^{kA})=\left(\begin{array}{ccc}\phi&0&0\\0&\psi&0\\0&0&\chi\end{array}\right),
\ee
where $k=1,2,3$, is the color index which runs vertically, $A=1,2,3$, is the
flavor index which runs horizontally, and the winding numbers of
$\phi,\psi,\chi$, away from a local region
where $\phi,\psi,\chi$ may vanish, say
$n_1,n_2,n_3$, characterize the quark fields, which will be identified as
the vortex charges or vortex numbers.

Now set
\be \label{A}
A_j^3=a_j,\quad A^8_j=\frac1{\sqrt{3}} b_j,\quad A_j=\frac13 c_j,\quad
j=1,2.
\ee
where $a_j,b_j,c_j$ ($j=1,2$) are real-valued vector fields.
Then, in terms of the fields given in (\ref{q}) and (\ref{A}), the
non-Abelian BPS multiple vortex equations (\ref{BPS1})--(\ref{BPS3}) are
found  \cite{Auzzi} to be
\bea
(\pa_1+\ii\pa_2)\phi&=&\ii\left(\frac12(a_1+\ii a_2)+\frac16(b_1+\ii
b_2)+\frac13(c_1+\ii c_2)\right)\phi,\label{1}\\
(\pa_1+\ii\pa_2)\psi&=&\ii\left(-\frac12(a_1+\ii a_2)+\frac16(b_1+\ii
b_2)+\frac13(c_1+\ii c_2)\right)\psi,\label{2}\\
(\pa_1+\ii\pa_2)\chi&=&\ii\left(-\frac13(b_1+\ii b_2)+\frac13(c_1+\ii
c_2)\right)\chi,\label{3}\\
a_{12}&=&-\frac\alpha2(|\phi|^2-|\psi|^2),\\
b_{12}&=&-\frac\alpha2(|\phi|^2+|\psi|^2-2|\chi|^2),\\
c_{12}&=&-\beta(|\phi|^2+|\psi|^2+|\chi|^2-3\xi),\label{6}
\eea
where
\be
a_{12}=\pa_1 a_2-\pa_2 a_1,\quad \mbox{etc},
\ee
are the reduced field curvatures, and $\alpha=g^2, \beta=3e^2$. For
convenience, we now use the complexified variables
\be\label{abc}
z=x^1+\ii x^2,\quad a=a_1+\ii a_2,\quad b=b_1+\ii b_2,\quad c=c_1+\ii c_2,
\ee
and the complex derivatives
\be
\pa=\pa_z=\frac12(\pa_1-\ii\pa_2),\quad
\overline{\pa}=\pa_{\overline{z}}=\frac12(\pa_1+\ii\pa_2)
\ee
to convert (\ref{1})--(\ref{6}) into the system
\bea
\overline{\pa}\ln\phi&=&\frac\ii2\left(\frac a2+\frac b6+\frac
c3\right),\label{7}\\
\overline{\pa}\ln\psi&=&\frac\ii2\left(-\frac a2+\frac b6+\frac
c3\right),\label{8}\\
\overline{\pa}\ln \chi&=&\frac\ii2\left(-\frac b3+\frac
c3\right),\label{9}\\
\ii(\pa
a-\overline{\pa}\overline{a})&=&\frac\alpha2(|\phi|^2-|\psi|^2),\label{10}\\
\ii(\pa
b-\overline{\pa}\overline{b})&=&\frac\alpha2(|\phi|^2+|\psi|^2-2|\chi|^2),\label{11}\\
\ii(\pa
c-\overline{\pa}\overline{c})&=&\beta(|\phi|^2+|\psi|^2+|\chi|^2-3\xi),\label{12}
\eea
away from the zeros of $\phi,\psi,\chi$. From (\ref{7})--(\ref{9}), we have
\bea
a&=&-2\ii\overline{\pa}(\ln\phi-\ln\psi),\label{13}\\
b&=&-2\ii\overline{\pa}(\ln\phi+\ln\psi-2\ln\chi),\label{14}\\
c&=&-2\ii\overline{\pa}(\ln\phi+\ln\psi+\ln\chi).\label{15}
\eea
Inserting (\ref{13})--(\ref{15}) into (\ref{10})--(\ref{12}) and using the
result $\Delta=4\pa\overline{\pa}=4\overline{\pa}\pa$, we obtain
\bea
\Delta(\ln|\phi|^2-\ln|\psi|^2)&=&\alpha(|\phi|^2-|\psi|^2),\label{16}\\
\Delta(\ln|\phi|^2+\ln|\psi|^2-2\ln|\chi|^2)&=&\alpha(|\phi|^2+|\psi|^2-2|\chi|^2),\label{17}\\
\Delta(\ln|\phi|^2+\ln|\psi|^2+\ln|\chi|^2)&=&2\beta(|\phi|^2+|\psi|^2+|\chi|^2-3\xi),\label{18}
\eea
again away from the zeros of $\phi,\psi,\chi$. Following \cite{JT}, we know
that the equations (\ref{1})--(\ref{3}) or (\ref{7})--(\ref{9}) imply that
the zeros of $\phi,\psi,\chi$ are discrete and of
integer multiplicities. We use $Z_\phi,Z_\psi,Z_\chi$ to denote the sets of
zeros of $\phi,\psi,\chi$,
\be\label{Z}
Z_\phi=\{p_{1,1},\cdots,p_{1,{n_1}}\},\quad
Z_\psi=\{p_{2,1},\cdots,p_{2,n_2}\},\quad
Z_\chi=\{p_{3,1},\cdots,p_{3,n_3}\},
\ee
so that the repetitions among the points $p_{\ell,s}$,
$\ell=1,2,3,s=1,\cdots,n_\ell$, take account of the multiplicities of these
zeros. Then the substitutions
\be
u_1=\ln|\phi|^2,\quad u_2=\ln|\psi|^2,\quad u_3=\ln|\chi|^2,
\ee
enable us to recast the equations (\ref{16})--(\ref{18}) into the following
elliptic system
\bea
\Delta(u_1-u_2)&=&\alpha
(\e^{u_1}-\e^{u_2})+4\pi\left(\sum_{s=1}^{n_1}\delta_{p_{1,s}}(x)-\sum_{s=1}^{n_2}\delta_{p_{2,s}}(x)\right),\label{19}\\
\Delta(u_1+u_2-2u_3)&=&\alpha(\e^{u_1}+\e^{u_2}-2\e^{u_3})\nn\\
&&\quad
+4\pi\left(\sum_{s=1}^{n_1}\delta_{p_{1,s}}(x)+\sum_{s=1}^{n_2}\delta_{p_{2,s}}(x)-2\sum_{s=1}^{n_3}\delta_{p_{3,s}}(x)\right),\,\,\label{20}\\
\Delta(u_1+u_2+u_3)&=&2\beta(\e^{u_1}+\e^{u_2}+\e^{u_3}-3\xi)\nn\\
&&\quad
+4\pi\left(\sum_{s=1}^{n_1}\delta_{p_{1,s}}(x)+\sum_{s=1}^{n_2}\delta_{p_{2,s}}(x)+\sum_{s=1}^{n_3}\delta_{p_{3,s}}(x)\right),\label{21}
\eea
now defined over the entire domain.

Two situations are of interest, namely, the situation where the equations
are considered over a doubly periodic domain, $\Om$, governing multiple
vortices hosted in $\Om$ so that the field
configurations are subject to the 't Hooft periodic boundary condition
\cite{tH0,WY,Ybook} under which periodicity is achieved modulo gauge
transformations, and the situation where the equations are considered
over the full plane $\bfR^2$ and the solutions satisfy the boundary
condition
\be
u_\ell(x)\to\ln\xi\quad\mbox{as }|x|\to\infty,\quad \ell=1,2,3.
\ee

Concerning these situations, our main existence and uniqueness theorem for
solutions of (\ref{1})--(\ref{6}) or (\ref{19})--(\ref{21}) may be stated as
follows.

\begin{theorem}
Consider the BPS system of multiple vortex equations (\ref{1})--(\ref{6})
for $(\phi,\psi,\chi,a_j,b_j,c_j)$ with the prescribed sets
of zeros given in (\ref{Z}) so that $\phi,\psi,\chi$ have $n_1, n_2, n_3$
arbitrarily distributed zeros, respectively.

(i) For this problem over a doubly periodic domain $\Om$, a solution exists
if and only if the following three conditions
\bea
\frac1{3\alpha}(n_1+n_2-2n_3)+\frac1\alpha(n_1-n_2)+\frac1{3\beta}(n_1+n_2+n_3)&<&\frac{\xi|\Om|}{2\pi},\label{24}\\
\frac1{3\alpha}(n_1+n_2-2n_3)-\frac1\alpha(n_1-n_2)+\frac1{3\beta}(n_1+n_2+n_3)&<&\frac{\xi|\Om|}{2\pi},\label{25}\\
-\frac2{3\alpha}(n_1+n_2-2n_3)+\frac1{3\beta}(n_1+n_2+n_3)&<&\frac{\xi|\Om|}{2\pi},\label{26}
\eea
hold simultaneously. Moreover, whenever a solution exists, it is unique.

(ii) For this problem over the full plane $\bfR^2$ subject to the boundary
condition
\be
|\phi|^2,|\psi|^2,|\chi|^2\to\xi\quad\mbox{as }|x|\to\infty,
\ee
there exists a unique solution up to gauge transformations so that the
boundary behavior stated above is realized exponentially rapidly.

In either case, the excited total vortex fluxes are quantized quantities
given explicitly by the formulas
\bea
\Phi_a=\int a_{12}\,{\rm \dd} x&=&2\pi(n_1-n_2),\\ \Phi_b=\int
b_{12}\,{\rm\dd} x&=&2\pi(n_1+n_2-2n_3),\\ \Phi_c=\int c_{12}\,{\rm\dd}
x&=&2\pi (n_1+n_2+n_3),
\eea
respectively
\end{theorem}

This theorem will be established in the following two sections.

\section{Proof of existence for doubly periodic case}
\setcounter{equation}{0}

In this section, we consider the equations (\ref{19})--(\ref{21}) defined
over a doubly periodic domain, $\Om$. Let $u_\ell^0$ be a solution of the
equation
\be
\Delta u_\ell^0=-\frac{4\pi n_\ell}{|\Om|}+4\pi\sum_{s=1}^{n_\ell}
\delta_{p_{\ell,s}}(x),\quad x\in\Om,\quad \ell=1,2,3.
\ee
Then the substitutions
\[
u_\ell= u_\ell^0+v_\ell,\quad\ell=1,2,3,
\]
recast the equations (\ref{19})--(\ref{21}) into
\bea
\Delta(v_1-v_2)&=&\alpha
(\e^{u^0_1+v_1}-\e^{u^0_2+v_2})+\frac{4\pi}{|\Om|}(n_1-n_2),\label{a2}\\
\Delta(v_1+v_2-2v_3)&=&\alpha(\e^{u^0_1+v_1}+\e^{u^0_2+v_2}-2\e^{u^0_3+v_3})+\frac{4\pi}{|\Om|}(n_1+n_2-2n_3),\label{a3}\\
\Delta(v_1+v_2+v_3)&=&2\beta(\e^{u^0_1+v_1}+\e^{u^0_2+v_2}+\e^{u^0_3+v_3}-3\xi)+\frac{4\pi}{|\Om|}(n_1+n_2+n_3).\,\,\label{a4}
\eea
Naturally, we should use the transformation
\be \label{T}
\left\{\begin{array}{lrl} w_1&=&v_1-v_2,\\w_2&=&v_1+v_2-2v_3,\\
w_3&=&v_1+v_2+v_3,\end{array}\right.\quad
\left\{\begin{array}{lrl}
v_1&=&\frac12w_1+\frac16w_2+\frac13w_3,\\v_2&=&-\frac12 w_1+\frac16
w_2+\frac13 w_3,\\ v_3&=&-\frac13 w_2+\frac13 w_3,\end{array}\right.
\ee
to change (\ref{a2})--(\ref{a4}) into the equations
\bea
\Delta w_1&=&\alpha
(\e^{u^0_1+\frac12w_1+\frac16w_2+\frac13w_3}-\e^{u^0_2-\frac12 w_1+\frac16
w_2+\frac13 w_3})+\frac{4\pi}{|\Om|}(n_1-n_2),\label{a6}\\
\Delta
w_2&=&\alpha(\e^{u^0_1+\frac12w_1+\frac16w_2+\frac13w_3}+\e^{u^0_2-\frac12
w_1+\frac16 w_2+\frac13 w_3}-2\e^{u^0_3-\frac13 w_2+\frac13 w_3})\nn\\
&&+\frac{4\pi}{|\Om|}(n_1+n_2-2n_3),\label{a7}\\
\Delta
w_3&=&2\beta(\e^{u^0_1+\frac12w_1+\frac16w_2+\frac13w_3}+\e^{u^0_2-\frac12
w_1+\frac16 w_2+\frac13 w_3}+\e^{u^0_3-\frac13 w_2+\frac13 w_3}-3\xi)\nn\\
&&+\frac{4\pi}{|\Om|}(n_1+n_2+n_3),\,\,\label{a8}
\eea
which are easily seen to be the Euler--Lagrange equations of the functional
\bea \label{I1st}
I(w_1,w_2,w_3)&=&\int_\Om\left\{\frac1{4\alpha}|\nabla
w_1|^2+\frac1{12\alpha}|\nabla w_2|^2+\frac1{12\beta}|\nabla
w_3|^2+\e^{u^0_1+\frac12w_1+\frac16w_2+\frac13w_3}\right.\nn\\
&&+\e^{u^0_2-\frac12 w_1+\frac16 w_2+\frac13 w_3}+\e^{u^0_3-\frac13
w_2+\frac13 w_3}+\frac{2\pi}{\alpha|\Om|}(n_1-n_2) w_1\nn\\
&&\left.+\frac{2\pi}{3\alpha|\Om|}(n_1+n_2-2n_3)
w_2+\left(\frac{2\pi}{3\beta|\Om|}(n_1+n_2+n_3)-\xi\right) w_3\right\}\,\dd
x.\nn\\
\eea
This functional is not bounded from below when
\be \label{ub}
2\pi (n_1+n_2+n_3)>3\beta\xi|\Om|.
\ee
In fact, we can show that (\ref{ub}) will never happen for
(\ref{a6})--(\ref{a8}). Indeed, integrating (\ref{a6})--(\ref{a8}), we
obtain the conditions
\bea
&&\int_\Om \e^{u_1^0+\frac12w_1+\frac16w_2+\frac13w_3}\,\dd x\nn\\
&&=\xi|\Om|
-2\pi\left(\frac1{3\alpha}(n_1+n_2-2n_3)+\frac1\alpha(n_1-n_2)+\frac1{3\beta}(n_1+n_2+n_3)\right)\nn\\
&&\equiv\eta_1>0,\label{cc1}\\
&&\int_\Om \e^{u^0_2-\frac12 w_1+\frac16 w_2+\frac13 w_3}\, \dd x\nn\\
&&=\xi|\Om|-2\pi\left(\frac1{3\alpha}(n_1+n_2-2n_3)-\frac1\alpha(n_1-n_2)+\frac1{3\beta}(n_1+n_2+n_3)\right)\nn\\
&&\equiv\eta_2>0,\label{cc2}\\
&&\int_\Om \e^{u^0_3-\frac13 w_2+\frac13 w_3}\,\dd x\nn\\
&&=\xi|\Om|
-2\pi\left(-\frac2{3\alpha}(n_1+n_2-2n_3)+\frac1{3\beta}(n_1+n_2+n_3)\right)\nn\\
&&\equiv\eta_3>0,\label{cc3}
\eea
which are exactly the conditions (\ref{24})--(\ref{26}). In particular, we
have
\be
\eta_1+\eta_2+\eta_3>0,
\ee
which rules out (\ref{ub}) immediately.

Below, we shall show that, under the conditions (\ref{cc1})--(\ref{cc3}),
the equations (\ref{a6})--(\ref{a8}) have a solution. We will use both
 a direct minimization method and a constrained minimization method to
approach the problem. These methods may be of independent practical value
for computational purposes.

\subsection{Direct minimization}
\newcommand{\ud}{\underline}

We use $W^{1,2}(\Om)$ to denote the usual Sobolev space of scalar-valued or
vector-valued $\Om$-periodic $L^2$-functions whose derivatives are also in
$L^2(\Om)$.
In the scalar case, we may decompose
$W^{1,2}(\Om)$ into $W^{1,2}(\Om)=\bfR\oplus \dot{W}^{1,2}(\Om)$ so that any
$f\in W^{1,2}(\Om)$ can be expressed as
\be \label{f}
f=\underline{f}+\dot{f},\quad\underline{f}\in\bfR,\quad \dot{f}\in
\dot{W}^{1,2}(\Om),\quad \int_\Om \dot{f}\,\dd x=0.
\ee

It is useful to recall the Moser--Trudinger inequality \cite{Aubin,Fon}
\be \label{MT}
\int_\Om \e^u\dd x\leq C\exp\left(\frac1{16\pi}\int_\Om |\nabla u|^2\,\dd
x\right),\quad u\in \dot{W}^{1,2}(\Om).
\ee

With (\ref{MT}), it is clear that the functional $I$ defined by (\ref{I1st})
is a $C^1$-functional with respect to its argument $(w_1,w_2,w_3)\in
W^{1,2}(\Om)$ which is strictly convex and
lower semi-continuous in terms of the weak topology of $W^{1,2}(\Om)$.

With the notation (\ref{f}), we may apply the transformation (\ref{T}) to
arrive at
\bea \label{Iv}
&&I(w_1,w_2,w_3)-\int_\Om\left\{\frac1{4\alpha}|\nabla
\dot{w}_1|^2+\frac1{12\alpha}|\nabla \dot{w}_2|^2+\frac1{12\beta}|\nabla
\dot{w}_3|^2\right\}\,\dd x\nn\\
&=&\int_\Om\left\{\e^{u^0_1+\ud{v}_1+\dot{v}_1}+\e^{u^0_2+\ud{v}_2+\dot{v}_2}+\e^{u^0_3+\ud{v}_3+\dot{v}_3}\right\}\,\dd
x-\eta_1\ud{v}_1-\eta_2\ud{v}_2-\eta_3\ud{v}_3\nn\\
&\geq&\sum_{\ell=1}^3\left(\sigma_\ell \e^{\ud{v}_\ell}-\eta_\ell
\ud{v}_\ell\right)\geq
\sum_{\ell=1}^3\eta_\ell\left(1+\ln\left[\frac{\sigma_\ell}{\eta_\ell}\right]\right),
\eea
where we have used the Jensen inequality to obtain the lower bounds
\bea \label{sigma}
\int_\Om \e^{u^0_\ell +\ud{v}_\ell+\dot{v}_\ell}\,\dd
x&\geq&|\Om|\exp\left(\frac1{|\Om|}\int_\Om (u^0_\ell
+\ud{v}_\ell+\dot{v}_\ell)\,\dd x\right)\nn\\
&=&\left(|\Om|\exp\left[\frac1{|\Om|}\int_\Om u^0_\ell\,\dd
x\right]\right)\e^{\ud{v}_\ell}\equiv\sigma_\ell\e^{\ud{v}_\ell},\quad
\ell=1,2,3,
\eea
in (\ref{Iv}). Thus, in particular, we see that $I$ is  bounded from below
and we may consider the following direct minimization problem
\be \label{min}
\eta_0\equiv \inf\left\{I(w_1,w_2,w_3)\bigg|\, w_1, w_2, w_3\in
W^{1,2}(\Om)\right\}.
\ee

Let $\{(w_1^{(n)},w_2^{(n)},w_3^{(n)})\}$ be a minimizing sequence of
(\ref{min}).
Since the function
\be\label{Fu}
F(u)=\sigma \e^u-\eta u,
\ee
 where $\sigma,\eta>0$ are constants, enjoys the property that
$F(u)\to\infty$ as $u\to\pm\infty$,
we see from (\ref{Iv}) that the sequences $\{\ud{v}_\ell^{(n)}\}$
($\ell=1,2,3$), hence  $\{\ud{w}_\ell^{(n)}\}$ ($\ell=1,2,3$), are all
bounded. Without loss of generality, we may assume
\be \label{2.20}
\ud{w}_\ell^{(n)}\to \mbox{ some point }\ud{w}^{(\infty)}_\ell\in \bfR\mbox{
as }n\to\infty,\quad \ell=1,2,3.
\ee

On the other hand, in view of (\ref{Iv}) and the Poincar\'{e} inequality, we
see that all the sequences $\{\dot{w}^{(n)}_\ell\}$ are bounded in
$\dot{W}^{1,2}(\Om)$, $\ell=1,2,3$.
Without loss of generality, we may assume
\be \label{2.21}
\dot{w}^{(n)}_\ell\to \mbox{ some element } \dot{w}^{(\infty)}_\ell\in
W^{1,2}(\Om) \mbox{ weakly as }n\to\infty,\quad \ell=1,2,3.
\ee

Of course, $\dot{w}^{(\infty)}\in \dot{W}^{1,2}(\Om)$ ($\ell=1,2,3$). Set
$w^{(\infty)}_\ell =\ud{w}^{(\infty)}_\ell+\dot{w}^{(\infty)}_\ell$
($\ell=1,2,3$). Then
(\ref{2.20}) and (\ref{2.21}) lead us to $w^{(n)}\to w^{(\infty)}$ weakly in
$W^{1,2}(\Om)$ as $n\to\infty$ ($\ell=1,2,3$). The weakly lower
semi-continuity of $I$ enables
us to conclude that $(w^{(\infty)}_1,w^{(\infty)}_2,w^{(\infty)}_3)$ solves
(\ref{min}), which is a critical point of $I$. As a critical point of $I$,
it satisfies the equations
(\ref{a6})--(\ref{a8}). Since $I$ is strictly convex, it can have at most
one critical point. Thus, the uniqueness of the solution of
(\ref{a6})--(\ref{a8}) follows immediately.

\subsection{Constrained minimization}

For convenience, we rewrite the constraints (\ref{cc1})--(\ref{cc3})
collectively as
\be \label{cc4}
J_\ell(w_1,w_2,w_3)\equiv \int_\Om\e^{u^0_\ell+{v}_\ell}\,\dd
x=\eta_\ell,\quad\ell=1,2,3,
\ee
and consider the constrained minimization problem
\be\label{cmin}
\eta_0\equiv \inf\left\{I(w_1,w_2,w_3)\bigg|\, (w_1, w_2, w_3)\in
W^{1,2}(\Om)\mbox{ and satisfies (\ref{cc4})}\right\}.
\ee

Suppose that (\ref{cmin}) allows a solution, say
$(\tilde{w}_1,\tilde{w}_2,\tilde{w}_3)$. Then there are numbers (the
Lagrange multipliers)
in $\bfR$, say $\lm_1,\lm_2,\lm_3$, such that
\bea \label{L}
&&D(I+\lm_1J_1+\lm_2 J_2+\lm_3
J_3)(\tilde{w}_1,\tilde{w}_2,\tilde{w}_3)(w_1,w_2,w_3)=0,\nn\\
&&\quad\forall (w_1,w_2,w_3)\in W^{1,2}(\Om).
\eea
Now take the trial configurations, $(w_1,w_2,w_3)=(1,0,0), (0,1,0), (0,0,1)$
consecutively, in (\ref{L}). Since $(\tilde{w}_1,\tilde{w}_2,\tilde{w}_3)$
satisfies the
constraints (\ref{cc4}), as a result, we have
\be
\frac12\lm_1\eta_1-\frac12\lm_2\eta_2=0,\quad
\frac16\lm_1\eta_1+\frac16\lm_2\eta_2-\frac13\lm_3\eta_3=0,\quad
\frac13\lm_1\eta_1+\frac13\lm_2\eta_2+\frac13\lm_3\eta_3=0.
\ee
Consequently, $\lm_1=\lm_2=\lm_3=0$. In other words, the Lagrange
multipliers disappear automatically and, thus, the search for a critical
point of the functional $I$ is converted
to obtaining a solution of the constrained minimization problem
(\ref{cmin}).

In order to approach (\ref{cmin}), we resolve (\ref{cc4}) to write down
\be \label{cc5}
\ud{v}_\ell =\ln\eta_\ell-\ln\left(\int_\Om\e^{u^0_\ell+\dot{v}_\ell}\,\dd
x\right),\quad\ell=1,2,3.
\ee
Hence, in view of the left-hand side of (\ref{Iv}), we get
\bea \label{Ivv}
&&I(w_1,w_2,w_3)-\int_\Om\left\{\frac1{4\alpha}|\nabla
\dot{w}_1|^2+\frac1{12\alpha}|\nabla \dot{w}_2|^2+\frac1{12\beta}|\nabla
\dot{w}_3|^2\right\}\,\dd x\nn\\
&\geq&-\sum_{\ell=1}^3\eta_\ell\ln\eta_\ell+\sum_{\ell=1}^3\eta_\ell\ln\left(\int_\Om
\e^{u^0_\ell +\dot{v}_\ell}\,\dd x\right)\geq
\sum_{\ell=1}^3\eta_\ell\ln\left(\frac{\sigma_\ell}{\eta_\ell}\right),
\eea
where we have used the Jensen inequality and the definition of the
quantities $\sigma_\ell$ ($\ell=1,2,3$) given in (\ref{sigma}). Thus the
problem (\ref{cmin}) is well defined.

Let $\{(w^{(n)}_1,w^{(n)}_2,w^{(n)}_3)\}$ be a minimizing sequence of
(\ref{cmin}). Then (\ref{Ivv}) says that the sequence
$\{(\dot{w}^{(n)}_1,\dot{w}^{(n)}_2,\dot{w}^{(n)}_3)\}$ is bounded in
$W^{1,2}(\Om)$. Hence we may assume that
$\{(\dot{w}^{(n)}_1,\dot{w}^{(n)}_2,\dot{w}^{(n)}_3)\}$ is weakly convergent
in $W^{1,2}(\Om)$. The inequality (\ref{MT}) and the expressions
(\ref{cc5}) indicate that
$\{(\ud{w}^{(n)}_1,\ud{w}^{(n)}_2,\ud{w}^{(n)}_3)\}$ is a convergent
sequence in $\bfR^3$. Thus $\{(w^{(n)}_1,w^{(n)}_2,w^{(n)}_3)\}$ is weakly
convergent in $W^{1,2}(\Om)$. In view of the weak continuity of the
constraint functionals $J_\ell$ defined in (\ref{cc4}) and the weak lower
semi-continuity of the functional $I$ defined in
(\ref{I}), we see that the weak limit of
$\{(w^{(n)}_1,w^{(n)}_2,w^{(n)}_3)\}$ in $W^{1,2}(\Om)$ is a solution of
(\ref{cmin}). As a critical of $I$, it is also unique.
Therefore, a constrained minimization proof for the existence of a unique
solution of the equations (\ref{a6})--(\ref{a8}) is obtained.

\section{Proof of existence for planar case}
\setcounter{equation}{0}

With the correspondence relations stated in (\ref{T}), we have
\bea
3w_1^2+w_2^2+2 w_3^2&=&6( v_1^2+ v_2^2+ v_3^2),\label{vw1}\\
3|\nabla w_1|^2+|\nabla w_2|^2+2|\nabla w_3|^2&=&6(|\nabla v_1|^2+|\nabla
v_2|^2+|\nabla v_3|^2),\label{vw2}
\eea
which will be useful for our analysis to follow.

To proceed further, here and elsewhere in the paper when we deal with the
planar cases, we use the method developed in \cite{JT} and introduce the
background functions  \cite{JT}
\be\label{u0}
u^0_\ell(x)=-\sum_{s=1}^{n_\ell}\ln(1+\mu|x-p_{\ell,s}|^{-2}),\quad
\mu>0,\quad\ell=1,2,3.
\ee
(Here and in the sequel, the parameter $\mu>0$ should not be confused with
the spacetime index $\mu$ used in the context of various field equations.)

Then we have
\be \label{h}
\Delta u^0_\ell=-h_\ell+4\pi\sum_{s=1}^{n_\ell}\delta_{p_{\ell,s}}(x),\quad
h_\ell(x)=4\sum_{s=1}^{n_\ell}\frac\mu{(\mu+|x-p_{\ell,s}|^2)^2},\quad
\ell=1,2,3.
\ee

Now use the substitutions
\bea
u_\ell&=&\ln\xi+u^0_\ell+v_\ell,\quad \ell=1,2,3,\\
\alpha\xi&\mapsto&\alpha,\quad \beta\xi\mapsto \beta,
\eea
and (\ref{T}) in (\ref{19})--(\ref{21}), we obtain the governing equations
\bea
\Delta w_1&=&\alpha
\left(\e^{u^0_1+\frac12w_1+\frac16w_2+\frac13w_3}-\e^{u^0_2-\frac12
w_1+\frac16 w_2+\frac13 w_3}\right)+(h_1-h_2),\label{3.6}\\
\Delta
w_2&=&\alpha\left(\e^{u^0_1+\frac12w_1+\frac16w_2+\frac13w_3}+\e^{u^0_2-\frac12
w_1+\frac16 w_2+\frac13 w_3}-2\e^{u^0_3-\frac13 w_2+\frac13 w_3}\right)\nn\\
&&+(h_1+h_2-2h_3),\label{3.7}\\
\Delta
w_3&=&2\beta\left(\e^{u^0_1+\frac12w_1+\frac16w_2+\frac13w_3}+\e^{u^0_2-\frac12
w_1+\frac16 w_2+\frac13 w_3}+\e^{u^0_3-\frac13 w_2+\frac13
w_3}-3\right)\nn\\
&&+(h_1+h_2+h_3),\,\,\label{3.8}
\eea
over the full plane $\bfR^2$. The boundary condition for $w_1,w_2,w_3$ reads
\be \label{bc}
w_\ell(x)\to0\quad\mbox{as }|x|\to\infty,\quad\ell=1,2,3.
\ee

In order to obtain a solution of (\ref{3.6})--(\ref{3.8}) subject to the
boundary condition (\ref{bc}), we look for a critical point of the action
functional
\bea \label{I}
I(w_1,w_2,w_3)&=&\int_{\bfR^2}\left\{\frac1{4\alpha}|\nabla
w_1|^2+\frac1{12\alpha}|\nabla w_2|^2+\frac1{12\beta}|\nabla
w_3|^2\right.\nn\\
&&+\left(\e^{u^0_1+\frac12w_1+\frac16w_2+\frac13w_3}-\e^{u^0_1}-\left[\frac12w_1+\frac16w_2+\frac13w_3\right]\right)\nn\\
&&+\left(\e^{u^0_2-\frac12 w_1+\frac16 w_2+\frac13
w_3}-\e^{u^0_2}-\left[-\frac12 w_1+\frac16 w_2+\frac13
w_3\right]\right)\nn\\
&&+\left(\e^{u^0_3-\frac13 w_2+\frac13 w_3}-\e^{u^0_3}-\left[-\frac13
w_2+\frac13 w_3\right]\right)\nn\\
&&\left.+\frac{1}{2\alpha}(h_1-h_2) w_1
+\frac{1}{6\alpha}(h_1+h_2-2h_3)
w_2+\frac{1}{6\beta}(h_1+h_2+h_3)w_3\right\}\,\dd x,\nn\\
\eea
which is $C^1$ and strictly convex over $W^{1,2}(\bfR^2)$. After some
algebra, it can be seen that the  Fr\'{e}chet derivative of
$I$ enjoys the following property,
\bea\label{DI1}
&&(DI(w_1,w_2,w_3))(w_1,w_2,w_3)-\int_{\bfR^2}\left\{\frac1{2\alpha}|\nabla
w_1|^2+\frac1{6\alpha}|\nabla w_2|^2+\frac1{6\beta}|\nabla
w_3|^2\right\}\,\dd x\nn\\
&&\quad\quad =\int_{\bfR^2}\left\{\sum_{\ell=1}^3
\e^{u_\ell^0}(\e^{v_\ell}-1)v_\ell+\sum_{\ell=1}^3 (\e^{u^0_\ell}-1)v_\ell
+ \sum_{\ell=1}^3 g_\ell v_\ell\right\}\,\dd x,
\eea
where $g_\ell$'s are some linear combinations of $h_\ell$'s. Now  setting
$\gamma=\max\{\alpha,2\beta\}$ and applying (\ref{vw2}), we derive from
(\ref{DI1}) the lower bound
\bea \label{3.13}
&&(DI(w_1,w_2,w_3))(w_1,w_2,w_3)\nn\\
&&\geq\int_{\bfR^2}\left\{\frac1{\gamma}\sum_{\ell=1}^3|\nabla
v_\ell|^2+\sum_{\ell=1}^3\left(\e^{u_\ell^0}[\e^{v_\ell}-1]v_\ell+
[\e^{u^0_\ell}-1]v_\ell  + g_\ell v_\ell\right)\right\}\,\dd x\nn\\
&&=\int_{\bfR^2}\left\{\frac1{\gamma}\sum_{\ell=1}^3|\nabla
v_\ell|^2+\sum_{\ell=1}^3 v_\ell\left(\e^{u_\ell^0+v_\ell}-1 + g_\ell
\right)\right\}\,\dd x.
\eea

We can now follow the analysis in \cite{JT}. To simplify the notation, we
suppress the subscript $\ell$ and rewrite a typical part on the right-hand
side of (\ref{3.13}) as
\be
M(v)=\int_{\bfR^2} v \left(\e^{u^0+v}-1 + g \right)\,\dd x.
\ee
Here $g$ should not be confused with the coupling constant used before in
the field-theoretical context. Thus $\e^t-1\geq t$ ($t\in\bfR$) gives us
\[
\e^{u^0+v}-1 + g\geq u^0+v+g,
\]
which leads to
\bea \label{3.15}
M(v_+)&\geq&\int_{\bfR^2}v^2_+\,\dd x+\int_{\bfR^2} v_+(u^0+g)\,\dd x\nn\\
&\geq&\frac12\int_{\bfR^2} v^2_+\,\dd x-\frac12\int_{\bfR^2} (u^0+g)^2\,\dd
x.
\eea
On the other hand, in view of the inequality $1-\e^{-t}\geq t/(1+t)$
($t\geq0$), we have
\bea \label{3.16}
v_-\left(1-g-\e^{u^0-v_-}\right)&=&v_-\left(1-g+\e^{u^0}[1-\e^{-v_-}]-\e^{u^0}\right)\nn\\
&\geq&v_-\left(1-g+\e^{u^0}\frac{v_-}{1+v_-}-\e^{u^0}\right)\nn\\
&=&\frac{v_-^2}{1+v_-}(1-g)+\frac{v_-}{1+v_-}\left(1-\e^{u^0}-g\right).
\eea
Of course, we may choose $\mu>0$ in (\ref{u0}) large enough so that $g<1/2$
(say). Furthermore, since $1-\e^{u^0}$ and $g$ are in $L^2(\bfR^2)$, we have
\be \label{3.17}
\int_{\bfR^2}\frac{v_-}{1+v_-}\left|1-\e^{u^0}-g\right|\,\dd
x\leq\frac14\int_{\bfR^2}\frac{v_-^2}{1+v_-}\,\dd
x+\int_{\bfR^2}\left(1-\e^{u^0}-g\right)^2\,\dd x.
\ee
Combining (\ref{3.16}) and (\ref{3.17}), we obtain
\be\label{3.18}
M(-v_-)=\int_{\bfR^2}v_-\left(1-g-\e^{u^0-v_-}\right)\,\dd x\nn\\
\geq\frac14\int_{\bfR^2}\frac{v_-^2}{1+v_-}\,\dd x-C,
\ee
where and in the sequel $C>0$ denotes an irrelevant constant. Summarizing
(\ref{3.15}) and (\ref{3.18}), we arrive at
\be \label{3.19}
M(v)\geq\frac14\int_{\bfR^2}\frac{v^2}{1+|v|}\,\dd x-C.
\ee

We now recall the standard Sobolev inequality
\be\label{3.20}
\int_{\bfR^2} v^4\,\dd x\leq 2\int_{\bfR^2} v^2\,\dd x\int_{\bfR^2}|\nabla
v|^2\,\dd x,\quad v\in W^{1,2}(\bfR^2).
\ee
Consequently, we have
\bea \label{3.21}
\left(\int_{\bfR^2}v^2\,\dd
x\right)^2&=&\left(\int_{\bfR^2}\frac{|v|}{1+|v|}(1+|v|)|v|\,\dd
x\right)^2\nn\\
&\leq&2\int_{\bfR^2}\frac{v^2}{(1+|v|)^2}\,\dd x\int_{\bfR^2}(v^2+v^4)\,\dd
x\nn\\
&\leq&4\int_{\bfR^2}\frac{v^2}{(1+|v|)^2}\,\dd x\int_{\bfR^2}v^2\,\dd
x\left(1+\int_{\bfR^2}|\nabla v|^2\,\dd x\right)\nn\\
&\leq&\frac12\left(\int_{\bfR^2}v^2\,\dd
x\right)^2+C\left(1+\left[\int_{\bfR^2}\frac{v^2}{(1+|v|)^2}\,\dd
x\right]^4+\left[\int_{\bfR^2}|\nabla v|^2\,\dd x\right]^4\right).\nn\\
&&
\eea
As a result of (\ref{3.21}), we have
\be \label{3.22}
\left(\int_{\bfR^2} v^2\,\dd x\right)^{\frac12}\leq
C\left(1+\int_{\bfR^2}|\nabla v|^2\,\dd
x+\int_{\bfR^2}\frac{v^2}{(1+|v|)^2}\,\dd x\right).
\ee

Now set $C_0=\min\{1/\gamma,1/4\}$. In view of (\ref{3.13}) and
(\ref{3.19}), we have
\be \label{3.23}
(DI(w_1,w_2,w_3))(w_1,w_2,w_3)\geq C_0 \sum_{\ell=1}^3
\int_{\bfR^2}\left(|\nabla v_\ell|^2+\frac{v_\ell^2}{1+|v_\ell|}\right)\,\dd
x -C.
\ee

As a consequence of (\ref{3.22}), (\ref{3.23}), (\ref{vw1}), (\ref{vw2}), we
may conclude with the coercive lower bound
\be \label{DI}
(DI(w_1,w_2,w_3))(w_1,w_2,w_3)\geq
C_1\sum_{\ell=1}^3\|w_\ell\|_{W^{1,2}(\bfR^2)} -C_2,
\ee
where $C_1,C_2>0$ are constants. In view of the estimate (\ref{DI}), the
existence of a critical point of the functional $I$ in the space
$W^{1,2}(\bfR^2)$ follows.
In fact, from (\ref{DI}), we may choose $R>0$ large enough so that
\be \label{DIR}
\inf\left\{ (DI(w_1,w_2,w_3))(w_1,w_2,w_3)\,\bigg|\,
\sum_{\ell=1}^3\|w_\ell\|_{W^{1,2}(\bfR^2)}=R\right\}\geq1
\ee
(say).  Consider the minimization problem
\be\label{Iinf}
\eta_0\equiv\inf\left\{
I(w_1,w_2,w_3)\,\bigg|\,\sum_{\ell=1}^3\|w_\ell\|_{W^{1,2}(\bfR^2)}\leq
R\right\}.
\ee
This problem obviously has a solution due to the fact that the functional
(\ref{I}) is weakly lower semi-continuous. Let
$(\tilde{w}_1,\tilde{w}_2,\tilde{w}_3)$ be a solution of
(\ref{Iinf}). We show that it must be an interior point. Otherwise, if
\be
\sum_{\ell=1}^3\|\tilde{w}_\ell\|_{W^{1,2}(\bfR^2)}=R,
\ee
then, with the vector notation ${\bf w}=(w_1,w_2,w_3)$, the result
(\ref{DIR}) gives us
\be
\lim_{t\to0}\frac{I([1-t]\tilde{\bf w})-I(\tilde{\bf w})}t=\frac{\dd}{\dd t}
I([1-t]\tilde{\bf w})\bigg|_{t=0}=-(DI(\tilde{\bf w}))(\tilde{\bf w})\leq
-1.
\ee
Thus, when $t>0$ is sufficiently small, with ${\bf w}^t=(1-t)\tilde{\bf w}$,
we have
\be
I(w_1^t,w_2^t,w_3^t)<I(\tilde{w}_1,\tilde{w}_2,\tilde{w}_3)=\eta_0,\quad
\sum_{\ell=1}^3\|w_\ell^t\|_{W^{1,2}(\bfR^2)}=(1-t)R<R,
\ee
which contradicts the definition of $\eta_0$ made in (\ref{Iinf}). Thus
$(\tilde{w}_1,\tilde{w}_2,\tilde{w}_3)$ must an interior point for the
problem (\ref{Iinf}). Consequently, it is
a critical point of the functional (\ref{I}).
The strict convexity of the functional implies that such a critical point
must be unique. In the following, we rewrite  $\tilde{w}_\ell$ as $w_\ell$.

Besides, using the standard embedding inequality
\be
\|f\|_{L^p(\bfR^2)}\leq \left(\pi\left[\frac
p2-1\right]\right)^{\frac{p-2}{2p}}\|f\|_{W^{1,2}(\bfR^2)},\quad p>2,
\ee
and the MacLaurin series
\be
(\e^f-1)^2=f^2+\sum_{s=3}^\infty \frac{2^s-2}{s!} f^s,
\ee
it is seen that $\e^f-1\in L^2(\bfR^2)$ when $f\in W^{1,2}(\bfR^2)$.
 Applying this in (\ref{3.6})--(\ref{3.8}) and using elliptic estimates,
we have $w_\ell\in W^{2,2}(\bfR^2)$ ($\ell=1,2,3$). In particular,
$w_\ell(x)\to0$ as $|x|\to\infty$, $\ell=1,2,3$. In view of this property
and (\ref{3.6})--(\ref{3.8}), we see that
the right-hand sides of (\ref{3.6})--(\ref{3.8}) all lie in $L^p(\bfR^2)$
for any $p>2$, which establishes $w_\ell\in W^{2,p}(\bfR^2)$ ($\ell=1,2,3$)
by elliptic $L^p$-estimates. Consequently,
$|\nabla w_\ell|(x)\to0$ as $|x|\to\infty$ ($\ell=1,2,3$). Linearizing
(\ref{3.6})--(\ref{3.8}), we see that $w_\ell$ vanishes exponentially fast
and $\nabla w_\ell$ vanishes
like $\mbox{O}(|x|^{-3})$  at infinity, $\ell=1,2,3$. Thus, we have
\be \label{3.25}
\int_{\bfR^2}\Delta w_\ell \,\dd x=0,\quad \ell=1,2,3.
\ee
Integrating  (\ref{3.6})--(\ref{3.8}) over $\bfR^2$ and inserting
(\ref{3.25}) and the definitions of $h_\ell$ ($\ell=1,2,3$), we have
\bea
&&\alpha
\int_{\bfR^2}\left(\e^{u^0_1+\frac12w_1+\frac16w_2+\frac13w_3}-\e^{u^0_2-\frac12
w_1+\frac16 w_2+\frac13 w_3}\right)\,\dd x\nn\\
&&=-\int_{\bfR^2}(h_1-h_2)\,\dd x
=-4\pi(n_1-n_2),\label{3.26}\\
&&\alpha\int_{\bfR^2}\left(\e^{u^0_1+\frac12w_1+\frac16w_2+\frac13w_3}+\e^{u^0_2-\frac12
w_1+\frac16 w_2+\frac13 w_3}-2\e^{u^0_3-\frac13 w_2+\frac13 w_3}\right)\,\dd
x\nn\\
&&=
-\int_{\bfR^2}(h_1+h_2-2h_3)\,\dd x
=-4\pi (n_1+n_2-2n_3),\label{3.27}\\
&&2\beta\int_{\bfR^2}\left(\e^{u^0_1+\frac12w_1+\frac16w_2+\frac13w_3}+\e^{u^0_2-\frac12
w_1+\frac16 w_2+\frac13 w_3}+\e^{u^0_3-\frac13 w_2+\frac13
w_3}-3\right)\,\dd x\nn\\
&&=
-\int_{\bfR^2}(h_1+h_2+h_3)\,\dd x=-4\pi(n_1+n_2+n_3),\,\,\label{3.28}
\eea
as stated in the theorem.

\section{Vortices in Chern--Simons--Higgs theory}
\setcounter{equation}{0}
In the context of supersymmetric Chern--Simons--Higgs theory in the standard
$(2+1)$-dimensional Minkowski spacetime, recently developed by Aharony,
Bergman,  Jafferis, and Maldacena \cite{ABJM},
known also as the ABJM model \cite{AK,Benna,GGRV,G,S,TY}, which is a
Chern--Simons theory within which the matter fields are four complex
scalars,
\be
C^I=(Q^1,Q^2,R^1,R^2),\quad I=1,2,3,4,
\ee
in the bi-fundamental $({\bf N},\overline{\bf N})$ representation
of the gauge group $U(N)\times U(N)$, which are all $N\times N$ complex
matrices, of the gauge fields $A_\mu$ and $ B_\mu$, and the associated
Chern--Simons terms for $A_\mu$
and $B_\mu$ are set at the levels $\kappa$ and $-\kappa$ so that they give
rise to the Lagrangian density
\be
{\cal L}_{\mbox{\tiny CS}}=\frac{\kappa}{4\pi}
\epsilon^{\mu\nu\gamma}\mbox{Tr}\left(A_\mu\pa_\nu A_\gamma+\frac{2\ii}3
A_\mu A_\nu A_\gamma -B_\mu\pa_\nu B_\gamma-\frac{2\ii}3 B_\mu B_\nu
B_\gamma\right),
\ee
and the gauge-covariant derivative
\be
D_\mu C^I=\pa_\mu C^I +\ii A_\mu C^I-\ii C^I B_\mu,\quad I=1,2,3,4.
\ee
The scalar potential density is of the mass-deformed form \cite{GGRV}
\be \label{V1}
V=\mbox{Tr} (M^{\alpha\dagger} M^\alpha+N^{\alpha\dagger} N^\alpha),
\ee
where
\bea
M^\alpha&=&\sigma Q^\alpha+\frac{2\pi}\kappa(2Q^{[\alpha}Q^\dagger_\beta
Q^{\beta]}+R^\beta R^\dagger_\beta Q^\alpha-Q^\alpha R^\dagger_\beta
R^\beta\nn\\
&&\quad +2Q^\beta R^\dagger_\beta R^\alpha-2R^\alpha R^\dagger_\beta
Q^\beta),\label{M}\\
N^\alpha&=&-\sigma R^\alpha+\frac{2\pi}\kappa(2R^{[\alpha}R^\dagger_\beta
R^{\beta]}+Q^\beta Q^\dagger_\beta R^\alpha-R^\alpha Q^\dagger_\beta
Q^\beta\nn\\
&&\quad +2R^\beta Q^\dagger_\beta Q^\alpha-2Q^\alpha Q^\dagger_\beta
R^\beta),\label{N}
\eea
the Kronecker symbol $\epsilon^{\alpha\beta}$ ($\alpha,\beta=1,2$) is used
to lower or raise indices, and $\sigma>0$ a massive parameter. Thus, when
the spacetime metric is of the signature
$(+--)$, the total (bosonic) Lagrangian density of the ABJM model can be
written as
\be \label{LABJM}
{\cal L}=-{\cal L}_{\mbox{\tiny CS}}+\mbox{Tr}([D_\mu C^I]^\dagger [D^\mu
C^I])-V,
\ee
which is of a pure Chern--Simons type for the gauge field sector. As in
\cite{AK}, we focus on a reduced situation where (say) $R^\alpha=0$. Then,
by virtue of (\ref{M}) and (\ref{N}), the scalar potential density
(\ref{V1}) takes the form
\be
V=\mbox{Tr} (M^{\alpha\dagger} M^\alpha),\quad M^\alpha=\sigma
Q^\alpha+\frac{4\pi}\kappa(Q^{\alpha}Q^\dagger_\beta
Q^{\beta}-Q^{\beta}Q^\dagger_\beta Q^{\alpha}).
\ee

The equations of motion of the Lagrangian (\ref{LABJM}) are rather
complicated. However, in the static limit, Auzzi and Kumar \cite{AK} showed
that these equations
may be reduced into the following first-order BPS system of equations
\bea
D_0 Q^1-\ii W^1&=&0,\quad D_1 Q^2+\ii D_2 Q^2=0,\\
D_1 Q^1&=&0, \quad D_2 Q^1=0,\quad D_0 Q^2=0, \quad W^2=0,
\eea
coupled with the Gauss law constraints which are the temporal components of
the Chern--Simons equations
\bea
\frac \kappa{4\pi}\epsilon^{\mu\nu\gamma} F^{(A)}_{\nu\gamma}&=&\ii (
Q^\alpha [D^\mu Q^\alpha]^\dagger-[D^\mu Q^\alpha]
Q^{\alpha\dagger}),\label{AF}\\
\frac \kappa{4\pi}\epsilon^{\mu\nu\gamma} F^{(B)}_{\nu\gamma}&=&\ii (
[D^\mu Q^\alpha]^\dagger Q^\alpha-Q^{\alpha\dagger}[D^\mu Q^\alpha]
),\label{BF}
\eea
where
\bea
F_{\mu\nu}^{(A)}&=&\pa_\mu A_\nu-\pa_\nu A_\mu +\ii[A_\mu,A_\nu],\quad \nn\\
F_{\mu\nu}^{(B)}&=&\pa_\mu B_\nu-\pa_\nu B_\mu +\ii[B_\mu,B_\nu],\nn\\
W^1&=&\sigma Q^1+\frac{2\pi}\kappa(Q^1 Q^{2\dagger}Q^2-Q^2
Q^{2\dagger}Q^1),\quad\nn\\
 W^2&=&\sigma Q^2+\frac{2\pi}\kappa(Q^2 Q^{1\dagger}Q^1-Q^1
Q^{1\dagger}Q^2),\nn
\eea
provided that \cite{AK} one takes the ansatz that $Q^1$ assumes its vacuum
expectation value
\be
Q^1=\sqrt{\frac{\sigma\kappa}{2\pi}}\,\mbox{diag}\left(0,1,\cdots,\sqrt{N-2},\sqrt{N-1}\right),
\ee
the non-trivial entries of $Q^2$ are given by $(N-1)$ complex scalar fields
$\psi$ and $\phi_\ell$ ($\ell=1,\cdots, N-2$) according to
\be
Q^2_{N,N-1}=\sqrt{\frac{\sigma\kappa}{2\pi}}\psi,\quad
Q^2_{N-\ell,N-\ell-1}=\sqrt{\frac{\sigma\kappa(\ell+1)}{2\pi}}\phi_\ell,\quad
\ell\nn=1,\cdots,N-2,
\ee
and the spatial components of the gauge fields $A_j$ and $ B_j$ ($j=1,2$)
are expressed in terms of $(N-1)$ real-valued vector potentials $a=(a_j))$
and
$b^\ell=(b^\ell_j)$ ($j=1,2; \ell=1,\cdots,N-2$) satisfying
\be
A_j=B_j=\mbox{diag}\left(0,b^{N-2}_j,\cdots, b^1_j, a_j\right),\quad j=1,2.
\ee

Within the above described formalism,  the non-Abelian BPS vortex equations
obtained by Auzzi and Kumar in \cite{AK}, without restricting to the
radially symmetric configurations, are of the form
\bea
(\pa_1+\ii\pa_2)\psi&=& \ii (a-b^1)\psi,\label{4.1}\\
(\pa_1+\ii\pa_2)\phi_\ell&=& \ii (b^\ell-b^{\ell+1})\phi_\ell,\quad 1\leq
\ell\leq N-3,\label{4.2}\\
(\pa_1+\ii\pa_2)\phi_{N-2}&=& \ii b^{N-2}\phi_{N-2},\label{4.3}\\
a_{12}&=&2(N-1)\sigma^2 (1-|\psi|^2),\label{4.4}\\
b^1_{12}&=&2(N-2)\sigma^2 (1+|\psi|^2-2|\phi_1|^2),\label{4.5}\\
b^\ell_{12}&=&2(N-1-\ell)\sigma^2
(1+\ell|\phi_{\ell-1}|^2-(\ell+1)|\phi_\ell|^2),\nn\\
 &&\quad 2\leq \ell\leq N-2,\label{4.6}
\eea
where
\[
a=(a_1,a_2)=a_1+\ii a_2, \quad b^\ell=(b_1^\ell,b_2^\ell)=b^\ell_1+\ii
b^\ell_2,\quad \ell=1,\cdots,N-2,
\]
 are the conveniently complexified gauge vector fields and the summation
convention is not applied to the repeated
index $\ell$. As before, the structure of the equations
(\ref{4.1})--(\ref{4.3}) implies that the zeros of the fields
$\psi,\phi_\ell$ ($\ell=1,\cdots,N-2$) are
discrete and of integer multiplicities which may collectively be expressed
in the form of the respective finite sets
\be \label{ZZ}
Z_\psi=\{p_{1,1},\cdots,p_{1,n_1}\},\quad
Z_{\phi_\ell}=\{p_{\ell+1,1},\cdots,p_{\ell+1,n_{\ell+1}}\},\quad
\ell=1,\cdots,N-2.
\ee

For the prescribed sets of zeros given in (\ref{ZZ}), we are to construct a
solution of (\ref{4.1})--(\ref{4.6}) to realize these zeros. For this
problem, here
is our main theorem.

\begin{theorem}
Consider the general BPS system of multiple vortex equations
(\ref{4.1})--(\ref{4.6}) for $(\psi,\phi_\ell,a,b^\ell)$ with the prescribed
sets
of zeros given in (\ref{ZZ}) so that $\psi,\phi_\ell$ have $n_1,
n_{\ell+1}$, $\ell=1,\cdots,N-2$, arbitrarily distributed zeros,
respectively.

(i) For this problem over a doubly periodic domain $\Om$, a solution exists
if and only if the following $(N-1)$ conditions
\bea
&&4\pi\sum_{k=1}^{N-1}n_k<(N-1)\lm|\Om|,\label{4.8} \quad \\
&&4\pi\left(\frac1{N-1}\sum_{k=1}^{N-1} n_k+\frac1{N-2}\sum_{k=2}^{N-1}
n_k\right)<2\lm|\Om|,\label{4.9}\\
&&4\pi\left(\frac1{N-1}\sum_{k=1}^{N-1} n_k+\frac1{N-2}\sum_{k=2}^{N-1}
n_k+\cdots+\frac1{N-\ell}\sum_{k=\ell}^{N-1} n_k\right)<\ell\lm|\Om|,\nn\\
&&\quad \ell=3,\cdots,N-1,\label{4.10}
\eea
are fulfilled simultaneously. Moreover, whenever a solution exists, it is
unique.

(ii) For this problem over the full plane $\bfR^2$ subject to the boundary
condition
\be
|\psi|^2,|\phi_\ell|^2\to 1\quad\mbox{as }|x|\to\infty, \quad
\ell=1,\cdots,N-2,
\ee
there exists a unique solution up to gauge transformations so that the
boundary behavior stated above is realized exponentially rapidly.

In either case, the excited total vortex fluxes are quantized quantities
given explicitly by the formulas
\bea
\Phi_a=\int a_{12}\,{\rm \dd} x&=&2\pi\sum_{k=1}^{N-1} n_k,\\
\Phi_{b^1}=\int b^1_{12}\,{\rm\dd} x&=&2\pi \sum_{k=2}^{N-1} n_k, \\
\Phi_{b^\ell}=\int b^\ell_{12}\,{\rm\dd} x&=&2\pi \sum_{k=\ell+1}^{N-1} n_k,
\quad \ell=2,\cdots,N-2,
\eea
respectively.
\end{theorem}

To approach the problem, we proceed as before.
Firstly, note that, away from the zero sets given in (\ref{ZZ}), we may
resolve the equations (\ref{4.1})--(\ref{4.3}) to find
\bea
a&=&-2\ii\overline{\pa}\left(\ln\psi+\ln\phi_1+\ln\phi_2+\cdots+\ln\phi_{N-2}\right),\label{4a}\\
b^1&=&-2\ii\overline{\pa}\left(\ln\phi_1+\ln\phi_2+\cdots+\ln\phi_{N-2}\right),\label{4b1}\\
b^\ell&=&-2\ii\overline{\pa}\left(\ln\phi_\ell+\cdots+\ln\phi_{N-2}\right),\quad
\ell=2,\cdots,N-3,\label{4bj}\\
b^{N-2}&=&-2\ii\overline{\pa}\ln\phi_{N-2}.\label{4bN}
\eea
Thus, following the same procedure as before to substitute
(\ref{4a})--(\ref{4bN}) into (\ref{4.4})--(\ref{4.6}), we arrive at
\bea
\Delta\left(\ln|\psi|^2+\ln|\phi_1|^2+\cdots+\ln|\phi_{N-2}|^2\right)&=&\lm
(N-1)(|\psi|^2-1),\label{4.19}\\
\Delta\left(\ln|\phi_1|^2+\cdots+\ln|\phi_{N-2}|^2\right)&=&\lm(N-2)(2|\phi_1|^2
-|\psi|^2-1),\label{4.20}\\
\Delta\left(\ln|\phi_\ell|^2+\cdots+\ln|\phi_{N-2}|^2\right)&=&\lm(N-1-\ell)([\ell+1]|\phi_\ell|^2
-\ell|\phi_{\ell-1}|^2-1),\nn\\
&&\quad \ell=2,\cdots,N-3,\label{4.21}\\
\Delta \ln|\phi_{N-2}|^2&=&\lm
([N-1]|\phi_{N-2}|^2-[N-2]|\phi_{N-3}|^2-1),\nn\\
\label{4.22}
\eea
away from the zero sets (\ref{ZZ}), where $\lm=4\mu^2$. Next, set $m=N-1$
and
\be
u_1=\ln|\psi|^2,\quad u_\ell=\ln|\phi_{\ell-1}|^2,\quad \ell=2,\cdots,N-1=m.
\ee
Then, the equations (\ref{4.19})--(\ref{4.22}) are converted into
\bea
\Delta(u_1+u_2+\cdots+u_m)&=&\lm
m\left(\e^{u_1}-1\right)+4\pi\sum_{k=1}^m\sum_{s=1}^{n_k}
\delta_{p_{k,s}}(x),\label{4.24}\\
\Delta(u_2+\cdots+ u_m)&=&\lm(m-1)\left(2\e^{u_2}-\e^{u_1}-1\right)\nn\\
&&\quad +4\pi\sum_{k=2}^m\sum_{s=1}^{n_k} \delta_{p_{k,s}}(x),\\
\Delta(u_\ell+\cdots+
u_m)&=&\lm(m-\ell+1)\left(\ell\e^{u_\ell}-[\ell-1]\e^{u_{\ell-1}}-1\right)\nn\\
&&\quad +4\pi\sum_{k=\ell}^m\sum_{s=1}^{n_k} \delta_{p_{k,s}}(x),
\quad \ell=3,\cdots,m-1,\\
\Delta u_m&=&\lm\left(m\e^{u_m}-[m-1]\e^{u_{m-1}}-1\right)\nn\\
&&\quad+4\pi\sum_{s=1}^{n_m}\delta_{p_{m,s}}(x).\label{4.27}
\eea

The above formalism may be viewed as an SQCD extension of those of Hong,
Kim, and Pac \cite{HKP}, Jackiw and Weinberg \cite{JW}, and Dunne
\cite{D1,D2}, of
the Abelian and non-Abelian Chern--Simons--Higgs theory.
See also the survey \cite{HZ}.

In the following two sections, we first consider the equations over a doubly
periodic domain. Then we consider the equations over the full plane.

\section{Proof of existence for doubly periodic case}
\setcounter{equation}{0}

Now consider the equations (\ref{4.24})--(\ref{4.27}) defined over a doubly
periodic domain, say $\Om$. We use the direct method to solve them.
Let $u_\ell^0$ be some doubly periodic source functions over $\Om$
satisfying
\be
\Delta u_\ell^0=-\frac{4\pi
n_\ell}{|\Om|}+4\pi\sum_{s=1}^{n_\ell}\delta_{p_{\ell,s}}(x),\quad
\ell=1,2,\cdots,m.
\ee
Then the substitutions $u_\ell=u^0_\ell+v_\ell$, $\ell=1,2,\cdots,m$, give
us the regularized equations
\bea
\Delta(v_1+v_2+\cdots+v_m)&=&\lm
m\left(\e^{u^0_1+v_1}-1\right)+\frac{4\pi}{|\Om|}\sum_{k=1}^m{n_k},\\
\Delta(v_2+\cdots+
v_m)&=&\lm(m-1)\left(2\e^{u^0_2+v_2}-\e^{u^0_1+v_1}-1\right)
+\frac{4\pi}{|\Om|}\sum_{k=2}^m{n_k},\\
\Delta(v_\ell+\cdots+
v_m)&=&\lm(m-\ell+1)\left(\ell\e^{u^0_\ell+v_\ell}-[\ell-1]\e^{u^0_{\ell-1}+v_{\ell-1}}-1\right)
+\frac{4\pi}{|\Om|}\sum_{k=\ell}^m{n_k},\nn\\
&&\quad j=3,\cdots,m-1,\\
\Delta
v_m&=&\lm\left(m\e^{u^0_m+v_m}-[m-1]\e^{u^0_{m-1}+v_{m-1}}-1\right)+\frac{4\pi}{|\Om|}{n_m}.
\eea

Integrating the above equations, we obtain the constraints
\bea
\int_\Om\e^{u_1^0+v_1}\,\dd
x&=&|\Om|-\frac{4\pi}{\lm}\left(\frac1m\sum_{k=1}^m
n_k\right)\equiv\eta_1>0,\label{ct1}\\
\int_\Om\e^{u_2^0+v_2}\,\dd
x&=&|\Om|-\frac{4\pi}{2\lm}\left(\frac1m\sum_{k=1}^m
n_k+\frac1{m-1}\sum_{k=2}^m n_k\right)\equiv\eta_2>0,\label{ct2}\\
\int_\Om\e^{u_\ell^0+v_\ell}\,\dd
x&=&|\Om|-\frac{4\pi}{\ell\lm}\left(\frac1m\sum_{k=1}^m
n_k+\frac1{m-1}\sum_{k=2}^m n_k+\cdots+\frac1{m-\ell+1}\sum_{k=\ell}^m
n_k\right)\nn\\
&\equiv&\eta_\ell>0,\quad \ell=3,\cdots,m,\label{ct3}
\eea
which are the conditions stated in (\ref{4.8})--(\ref{4.10}). Moreover, it
will be convenient to introduce the transformation
\be \label{TT}
\left\{\begin{array}{lll}
w_1&=&v_1+v_2+\cdots+v_m,\\
w_2&=& v_2+\cdots+v_m,\\
\cdots&=&\cdots\cdots\cdots\cdots\\
w_\ell&=& v_\ell+\cdots+v_m,\\
\cdots&=&\cdots\cdots\\
w_m&=&v_m,\end{array}\right.
\quad
\left\{\begin{array}{lll}
v_1&=&w_1-w_2,\\
v_2&=& w_2-w_3,\\
\cdots&=&\cdots\cdots\\
v_\ell&=& w_\ell-w_{\ell+1},\\
\cdots&=&\cdots\cdots\\
v_m&=&w_m.\end{array}\right.
\ee
Consequently, the governing equations become
\bea
\Delta w_1&=&
m\lm\left(\e^{u_1^0+w_1-w_2}-1\right)+\frac{4\pi}{|\Om|}\sum_{k=1}^m n_k,\\
\Delta w_2&=&
(m-1)\lm\left(2\e^{u_2^0+w_2-w_3}-\e^{u_1^0+w_1-w_2}-1\right)+\frac{4\pi}{|\Om|}\sum_{k=2}^m
n_k,\\
\Delta w_\ell&=&
(m-\ell+1)\lm\left(\ell\e^{u_\ell^0+w_\ell-w_{\ell+1}}-[\ell-1]\e^{u_{\ell-1}^0+w_{\ell-1}-w_\ell}-1\right)\nn\\
&&\quad+\frac{4\pi}{|\Om|}\sum_{k=\ell}^m n_k,
\quad \ell=3,\cdots,m-1,\\
\Delta w_m&=&
\lm\left(m\e^{u_m^0+w_m}-[m-1]\e^{u_{m-1}^0+w_{m-1}-w_m}-1\right)+\frac{4\pi}{|\Om|}
n_m,
\eea
whose variational functional is seen to be
\bea \label{II}
&&I(w_1,\cdots,w_m)\nn\\
&&=\int_\Om\left\{\frac1{2m\lm}|\nabla
w_1|^2+\cdots+\frac1{2(m-\ell+1)\lm}|\nabla
w_\ell|^2+\cdots+\frac1{2\lm}|\nabla w_m|^2\right\}\,\dd x\nn\\
&&\quad +J(w_1,\cdots,w_m),
\eea
where
\bea\label{JJ}
J(w_1,\cdots,w_m)&=&\int_\Om\left\{\left(\e^{u_1^0+w_1-w_2}-\left[1-\frac{4\pi}{m\lm|\Om|}\sum_{k=1}^m
n_k\right]w_1\right)\right.\nn\\
&&+\cdots+\left(\ell\e^{u_\ell^0+w_\ell-w_{\ell+1}}-\left[1-\frac{4\pi}{(m-\ell+1)\lm|\Om|}\sum_{k=\ell}^m
n_k\right]w_\ell\right)\nn\\
&&+\cdots+\left.\left(m\e^{u_m^0+w_m}-\left[1-\frac{4\pi}{\lm|\Om|}
n_m\right]w_m\right)\right\}\,\dd x.
\eea

On the other hand, in view of (\ref{TT}), we obtain after some algebra the
representation
\be \label{5.16}
J(w_1,\cdots,w_m)=\sum_{\ell=1}^m \ell \left(\int_\Om
\e^{u_\ell^0+v_\ell}\,\dd x-\eta_\ell \ud{v}_\ell\right).
\ee
Thus, we may use the same direct minimization method as before in a verbatim
way to establish the existence and uniqueness of a critical point of the
functional (\ref{II}).
\medskip
\medskip

For completeness, we now sketch how to establish the existence of a critical
point of (\ref{II}) by a constrained minimization approach. For this
purpose, we rewrite
(\ref{ct1})--(\ref{ct3}) as
\be \label{5.17}
J_\ell(w_1,\cdots,w_m)\equiv\int_\Om \e^{u_\ell^0+v_\ell}\,\dd
x=\eta_\ell,\quad \ell=1,\cdots,m,
\ee
and consider the problem
\be\label{5.18}
\min\left\{I(w_1,\cdots,w_m)\,|\,(w_1,\cdots,w_m)\mbox{ satisfies
(\ref{5.17}) and lies in }W^{1,2}(\Om)\right\}.
\ee

We use the notation ${\bf w}=(w_1,\cdots,w_m)$. If $\tilde{\bf w}$ is a
solution to (\ref{5.18}), then there are some numbers (the Lagrange
multipliers) $\lm_1,\cdots,\lm_m\in\bfR$ so that
\be \label{5.19}
\left(DI(\tilde{\bf w})+\lm_1 DJ_1(\tilde{\bf w})+\cdots+\lm_m
DJ_m(\tilde{\bf w})\right)({\bf w})=0,\quad\forall{\bf w}.
\ee
Now insert in (\ref{5.19}) the test configurations ${\bf w}={\bf
w}_\ell=(\delta_{1\ell},\cdots,\delta_{m\ell})$, $\ell=1,\cdots,m$. We have,
after applying (\ref{5.17}), the relations
\be
\lm_1\eta_1=0;\quad -\lm_{\ell-1}\eta_{\ell-1}+\lm_\ell\eta_\ell=0,\quad
\ell=2,\cdots,m,
\ee
which lead us to $\lm_1=\cdots=\lm_m=0$. In other words, the constraints do
not give rise to the undesired Lagrange multiplier problem so that a
solution of the constrained minimization
problem (\ref{5.18}) is a critical point of the functional (\ref{II})
itself.

Moreover, from the constraints (\ref{5.17}), we have
\bea
\ud{v}_\ell&=&\ud{w}_\ell-\ud{w}_{\ell+1}=\ln\eta_\ell-\ln\left(\int_\Om\e^{u_\ell^0+\dot{w}_\ell-\dot{w}_{\ell+1}}\,\dd
x\right),\quad \ell=1,\cdots,m-1,\\
\ud{v}_m&=&\ud{w}_m=\ln\eta_m-\ln\left(\int_\Om\e^{u_m^0+\dot{w}_m}\,\dd
x\right).
\eea

Inserting these into (\ref{5.16}) and applying the condition $\eta_\ell>0$
($\ell=1,\cdots,m$) and the Jensen inequality,  we again arrive at the
coerciveness
for the functional (\ref{II}),
\be
I(w_1,\cdots,w_m)\geq C_1\sum_{\ell=1}^m\int_\Om|\nabla \dot{w}_\ell|^2\,\dd
x-C_2,
\ee
where $C_1,C_2>0$ are some irrelevant constants. Consequently, the existence
of a solution to the problem (\ref{5.18}) follows as before.

\section{Proof of existence for planar case}
\setcounter{equation}{0}

To proceed, we define $u_\ell^0$ and $h_\ell$ as in (\ref{u0}) and (\ref{h})
so that $\ell$ runs from 1 through $m$. Thus, in view of the translations
$u_\ell=u_\ell^0+v_\ell$ ($\ell=1,\cdots,m$) and the transformation
(\ref{TT}), the governing equations (\ref{4.24})--(\ref{4.27}) become
\bea
\Delta w_1&=& m\lm\left(\e^{u_1^0+w_1-w_2}-1\right)+\sum_{k=1}^m h_k,\\
\Delta w_\ell&=&
(m-\ell+1)\lm\left(\ell\e^{u_\ell^0+w_\ell-w_{\ell+1}}-[\ell-1]\e^{u_{\ell-1}^0+w_{\ell-1}-w_\ell}-1\right)+\sum_{k=\ell}^m
h_k,\nn\\
&&\quad \ell=2,\cdots,m-1,\\
\Delta w_m&=&
\lm\left(m\e^{u_m^0+w_m}-[m-1]\e^{u_{m-1}^0+w_{m-1}-w_m}-1\right)+ h_m,
\eea
which are the Euler--Lagrange equations of the functional
\be\label{III}
I(w_1,\cdots,w_m)=\int_{\bfR^2}\left\{\frac1{2\lm}\sum_{\ell=1}^m\frac1{(m-\ell+1)}|\nabla
w_\ell|^2\right\}\,\dd x+J(w_1,\cdots,w_m),
\ee
where
\bea\label{6.5}
&&J(w_1,\cdots,w_m)=\nn\\
&&\int_{\bfR^2}\left\{\sum_{\ell=1}^{m-1}\left(\ell\e^{u_\ell^0+w_\ell-w_{\ell+1}}-\ell\e^{u_\ell^0}-w_\ell\right)+\left(m\e^{u_m^0+w_m}-m\e^{u_m^0}
-w_m\right)\right\}\,\dd x\nn\\
&&+\int_{\bfR^2}\left\{\frac1\lm\sum_{\ell=1}^{m-1}\frac1{(m-\ell+1)}\left(\sum_{k=\ell}^m
h_k\right) w_\ell+\frac1\lm h_m w_m\right\}\,\dd x.
\eea
It is important to note that, using the relation
\be
\sum_{\ell=1}^m w_\ell=\sum_{\ell=1}^{m-1} \ell(w_\ell-w_{\ell+1}) +m w_m,
\ee
we can rewrite $J(w_1,\cdots,w_m)$ defined in (\ref{6.5}) as
\bea\label{6.7}
&&J(w_1,\cdots,w_m)=\nn\\
&&\int_{\bfR^2}\left\{\sum_{\ell=1}^{m-1}\ell\left(\e^{u_\ell^0+w_\ell-w_{\ell+1}}-\e^{u_\ell^0}-[w_\ell-w_{\ell+1}]\right)+m\left(\e^{u_m^0+w_m}-\e^{u_m^0}
-w_m\right)\right\}\,\dd x\nn\\
&&+\int_{\bfR^2}\left\{\frac1\lm\sum_{\ell=1}^{m-1}\frac1{(m-\ell+1)}\left(\sum_{k=\ell}^m
h_k\right) w_\ell+\frac1\lm h_m w_m\right\}\,\dd x.
\eea
Consequently, after some algebraic manipulation, we obtain
\bea\label{6.8}
&&(DI(w_1,\cdots,w_m))(w_1,\cdots,w_m)=
\frac1{\lm}\int_{\bfR^2}\left\{\sum_{\ell=1}^m\frac1{(m-\ell+1)}|\nabla
w_\ell|^2\right\}\,\dd x\nn\\
&&+\int_{\bfR^2}\left\{\sum_{\ell=1}^{m-1}\left(\ell\e^{u_\ell^0}\left[\e^{w_\ell-w_{\ell+1}}-1\right][w_\ell-w_{\ell+1}]+\ell\left[\e^{u_\ell^0}-1\right][w_\ell-w_{\ell+1}]\right)\right.\nn\\
&&\quad+m\e^{u_m^0}\left[\e^{w_m}-1\right]w_m+m\left[\e^{u_m^0}-1\right]w_m\nn\\
&&\quad+\left.\frac1\lm\sum_{\ell=1}^{m-1}\frac1{(m-\ell+1)}\left(\sum_{k=\ell}^m
h_k\right) w_\ell+\frac1\lm h_m w_m\right\}\,\dd x.
\eea

On the other hand, in view of the transformation (\ref{TT}), we have
\bea
c_1\sum_{\ell=1}^m v_\ell^2&\leq& \sum_{\ell=1}^m w_\ell^2\leq
c_2\sum_{\ell=1}^m v_\ell^2,\label{6.9}\\
c_1\sum_{\ell=1}^m |\nabla v_\ell|^2&\leq& \sum_{\ell=1}^m |\nabla
w_\ell|^2\leq c_2\sum_{\ell=1}^m |\nabla v_\ell|^2,\label{6.10}
\eea
where $c_1,c_2>0$ are some constants. Thus (\ref{6.8}) and (\ref{6.10})
enable us to arrive at
\bea\label{6.11}
&&(DI(w_1,\cdots,w_m))(w_1,\cdots,w_m)\geq \nn\\
&&c_0\sum_{\ell=1}^m\int_{\bfR^2}|\nabla v_\ell|^2+\sum_{\ell=1}^m
\int_{\bfR^2}\left\{
\ell\left(\e^{u_\ell^0}\left[\e^{v_\ell}-1\right]v_\ell+\left[\e^{u_\ell^0}-1\right]v_\ell\right)+g_\ell
 v_\ell\right\}\,\dd x,\quad\quad\quad
\eea
where $c_0>0$ is a suitable constant and the functions $g_\ell$'s are some
linear combinations of the functions $h_\ell$'s. It has been seen that the
structure of the right-hand side of
(\ref{6.11}) indicates that there are constants $C_1,C_2>0$ such that
\be \label{6.12}
(DI(w_1,\cdots,w_m))(w_1,\cdots,w_m)\geq
C_1\sum_{\ell=1}^m\int_{\bfR^2}\left(|\nabla
v_\ell|^2+\frac{v_\ell^2}{1+|v_\ell|}\right)\,\dd x -C_2.
\ee
Therefore, applying (\ref{6.9}), (\ref{6.10}), and (\ref{6.12}), we can
again conclude with the coercive lower bound
\be
(DI(w_1,\cdots,w_m))(w_1,\cdots,w_m)\geq C_3\sum_{\ell=1}^m
\|w_\ell\|_{W^{1,2}(\bfR^2)}-C_4,
\ee
for some constants $C_3,C_4>0$. Hence, the existence of a critical point of
the functional (\ref{III}) in the space $W^{1,2}(\bfR^2)$ follows. Since
(\ref{III}) is strictly convex in
$(w_1,\cdots,w_m)\in W^{1,2}(\bfR^2)$ and $C^1$, it may have at most one
critical point in $W^{1,2}(\bfR^2)$.

The rest of the analysis regarding asymptotic estimates and computation of
fluxes is similar to that of Section 4 and is thus omitted.

\section{Further applications of direct methods}
\setcounter{equation}{0}

In this section, we show that our direct minimization methods may be used to
study other non-Abelian BPS vortex equations
of similar structures arising in SQCD. We will present two examples as
further illustrations.

\subsection{Vortices in an $SO(2N)$ theory}
In this subsection, we use the direct method developed earlier to strengthen
the existence results obtained in \cite{LY} for an $SO(2N)$ BPS vortex
problem formulated in \cite{GJK}.

Recall that, in the work of Gudnason--Jiang--Konishi \cite{GJK}, the
Lagrangian density of the non-Abelian Yang--Mills--Higgs model reads
\bea
{\cal L}&=&-\frac1{4e^2}F_{\mu\nu}^0 F^{0\mu\nu}-\frac1{4g^2}
F^a_{\mu\nu}F^{a\mu\nu}+({\cal D}_\mu q_f)^\dagger {\cal D}^\mu q_f\nn\\
&&\quad -\frac{e^2}2 \left| q_f^\dagger t^0
q_f-\frac{v_0^2}{\sqrt{4N}}\right|^2
-\frac{g^2}2 \left|q_f^\dagger t^a q_f \right|^2,
\eea
for which the gauge group $G$ is of the general form $G=G'\times U(1)$ where
$G'$ is a compact simple Lie group which may typically be chosen to be
$G'=SO(2N)$ or $G'=USp(2N)$ (the unitary
symplectic group). Assume that $a=1,\cdots,\dim(G')$ labels the generators
of $G'$, the index $0$ indicates
the $U(1)$ gauge field, $f=1,\cdots,N_{\mbox{\tiny flavor}}$ labels the
matter flavors or `scalar quark' fields, $q_f$, all are assumed to lie in
the fundamental representation of $G'$.
The gauge fields, gauge-covariant derivatives, and field tensors are given
by
\be
A_\mu=A^0_\mu+A_\mu^a t^a,\quad {\cal D}_\mu q_f=\pa_\mu q_f+\ii A_\mu
q_f,\quad F_{\mu\nu}=\pa_\mu A_\nu-\pa_\nu A_\mu+\ii[A_\mu,A_\nu],
\ee
respectively, where the generators of $G'$ and $U(1)$, i.e., $\{t^a\}$ and
$t^0$, are normalized to satisfy
\be
\mbox{Tr}(t^a t^b)=\frac12 \delta^{ab},\quad t^0=\frac1{\sqrt{4N}}{\bf
1}_{2N},
\ee
with ${\bf 1}_m$ denoting the $m\times m$ identity matrix. When the number
of matter flavors is $N_{\mbox{\tiny flavor}}=2N$, the scalar quark fields
may be represented as
a color-flavor mixed matrix $q$ of size $2N\times 2N$. Restricting to static
field configurations which are uniform with respect to the spatial
coordinate $x^3$, a Bogomol'nyi completion
\cite{Bo} may be performed to yield the BPS \cite{Bo,PS} vortex equations
\cite{Auzzi,EF,GJK,MY,ShY2004}
\bea
{\cal D}_1 q+\ii{\cal D}_2 q&=&0,\label{aBPS}\\
F^0_{12}-\frac{e^2}{\sqrt{4N}}(\mbox{Tr}(qq^\dagger)-v_0^2)&=&0,\label{bBPS}\\
F^a_{12}t^a-\frac{g^2}4(qq^\dagger-J^\dagger (qq^\dagger)^T
J)&=&0,\label{cBPS}
\eea
where $J$ is the standard symplectic matrix
\be
J=\left(\begin{array}{cc}0&{\bf 1}_N\\-{\bf 1}_N&0\end{array}\right).
\ee

In its general form, the system of the non-Abelian BPS vortex equations
(\ref{aBPS})--(\ref{cBPS}) appears hard to approach and an ansatz-based
reduction may be made as a tool for further
simplification. In the case when $G'=SO(2N)$, the ans\"{a}tze presented in
\cite{GJK} gives us the following
  matrix forms for the Higgs field,
\be
q=\left(\begin{array}{ccc}\Phi{\bf1}_{2N-2}&
0&0\\0&\phi&0\\0&0&\psi\end{array}\right),
\ee
where  $\Phi,\phi,\psi$ are three complex scalar fields, and for the gauge
potential,
\be
A_j=a_j{\bf1}_{2N}+b_j \mbox{diag}\{{\bf0}_{2N-2},1,-1\},\quad j=1,2,
\ee
where $a_j$ and $b_j$ ($j=1,2$) are real-valued vector fields.
Then, in terms of the complexified field $a$ and $b$ defined in (\ref{abc}),

 the BPS system of vortex equations found in \cite{GJK} assumes the form
\bea
\overline{\pa}\Phi&=&\ii a\Phi,\label{*BPS1}\\
\overline{\pa}\phi&=&\ii (a+b)\phi,\label{*BPS2}\\
\overline{\pa}\psi&=&\ii (a-b)\psi,\label{*BPS3}\\
a_{12}&=&\frac{e^2}{4N}(\xi^2-2(N-1)|\Phi|^2-|\phi|^2-|\psi|^2),\label{*BPS4}\\
b_{12}&=&\frac{g^2}4(|\psi|^2-|\phi|^2).\label{*BPS5}
\eea
Thus, we can recast the system of equations (\ref{*BPS1})--(\ref{*BPS5})
into
\bea
\Delta\ln|\Phi|^4&=&\Delta\left(\ln|\phi|^2+\ln|\psi|^2\right),\label{*4.8}\\
\Delta
\left(\ln|\phi|^2+\ln|\psi|^2\right)&=&\frac{2e^2}N\left(2(N-1)|\Phi|^2+|\phi|^2+|\psi|^2-\xi^2\right),\label{*4.9}\\
\Delta
\left(\ln|\phi|^2-\ln|\psi|^2\right)&=&2g^2(|\phi|^2-|\psi|^2),\label{*4.10}
\eea
where we have stayed away from the possible zeros of the fields $\Phi,
\phi,\psi$. We extend our study in \cite{LY} and consider a solution so that
the zeros of $\Phi$ coincide with those of $\phi$ and $\psi$. As a
consequence of the boundary condition, we see that
 (\ref{*4.8})  leads us to the simple relation
\be \label{*Phi}
|\Phi|^4=|\phi|^2|\psi|^2.
\ee

We are interested in constructing solutions over a doubly periodic domain,
$\Om$.
The multiple vortices are generated from the sets of zeros of $\phi$ and
$\psi$, prescribed as
\be \label{Zpq}
Z_\phi=\{p_1,\cdots,p_m\},\quad Z_\psi=\{q_1,\cdots,q_n\}.
\ee
Therefore the vortex-governing equations are then given in terms of the new
functions $u=\ln|\phi|^2$ and $v=\ln|\psi|^2$ as
\bea
\Delta
(u+v)&=&\alpha\left(2(N-1)\e^{\frac12(u+v)}+\e^u+\e^v-\gamma\right)+4\pi\sum_{s=1}^m\delta_{p_s}(x)+4\pi\sum_{s=1}^n\delta_{q_s}(x),\quad\quad
\quad\label{7.2}\\
\Delta(u-v)&=&\beta\left(\e^u-\e^v\right)+4\pi\sum_{s=1}^n\delta_{p_s}(x)-4\pi\sum_{s=1}^n\delta_{q_s}(x),\label{7.3}
\eea
where $\alpha,\beta,\gamma$ are positive constants given by
\be
\alpha=\frac{2e^2}N,\quad \beta=2g^2,\quad \gamma=\xi^2.
\ee
 In \cite{LY}, we proved an existence and uniqueness theorem for the
solution of (\ref{7.2}) and (\ref{7.3}) when $Z_\psi=\emptyset$
(or $n=0$) under the necessary and sufficient condition
\be\label{7.4}
4\pi m\left(\frac1\alpha+\frac1\beta\right)<\gamma|\Om|,
\ee
and we found suitable conditions under which the solution may be constructed
by a constrained minimization method. The general existence proof in
\cite{LY}, however, is based on
{\em a priori} estimates and a
degree theory argument which is unfortunately non-constructive. Here we show
that the solution can actually be obtained by the (constructive) direct
minimization method used in the earlier
sections of the present paper. For the broadest generality, we consider the
presence of the zeros of $\psi$ as well ($n\geq0$). We are able to obtain
the following sharp results.

\begin{theorem} For the non-Abelian vortex equations (\ref{7.2}) and
(\ref{7.3}) defined over the doubly periodic domain $\Om$, a solution exists
if and only if
the condition
\be\label{7.5}
4\pi\left(\frac{(m+n)}\alpha+\frac{|m-n|}\beta\right)<\gamma|\Om|.
\ee
Furthermore, if a solution exists, it is unique and may be constructed by a
direct minimization method.
\end{theorem}

To proceed, let $u_0$ and $v_0$ be solutions of the equations
\be\label{uv00}
\Delta u_0=4\pi\sum_{s=1}^m\delta_{p_s}(x)-\frac{4\pi m}{|\Om|},\quad \Delta
v_0=4\pi\sum_{s=1}^n\delta_{q_s}(x)-\frac{4\pi n}{|\Om|}.
\ee
Then the substitutions $u=u_0+U$ and $v=v_0+V$ change the equations
(\ref{7.2}) and (\ref{7.3}) into
\bea
\Delta
(U+V)&=&\alpha\left(2(N-1)\e^{\frac12(u_0+v_0)+\frac12(U+V)}+\e^{u_0+U}+\e^{v_0+V}-\gamma\right)\nn\\
&&\quad +\frac{4\pi}{|\Om|}(m+n),\\
\Delta(U-V)&=&\beta\left(\e^{u_0+U}-\e^{v_0+V}\right)+\frac{4\pi}{|\Om|}(m-n).
\eea
Next, use the transformation
\be
\left\{\begin{array}{rl}
f&=\frac12(U+V),\\g&=\frac12(U-V),\end{array}\right.\quad
\left\{\begin{array}{rl} U&=f+g,\\V&=f-g.\end{array}\right.
\ee
We see that $f,g$ satisfy
\bea
\Delta
f&=&\alpha\left((N-1)\e^{\frac12(u_0+v_0)+f}+\frac12\e^{u_0+f+g}+\frac12\e^{v_0+f-g}-\frac\gamma2\right)
+\frac{2\pi}{|\Om|}(m+n),\quad\quad\label{7f}\\
\Delta
g&=&\frac\beta2\left(\e^{u_0+f+g}-\e^{v_0+f-g}\right)+\frac{2\pi}{|\Om|}(m-n),\label{7g}
\eea
which are the Euler--Lagrange equations of the functional
\bea\label{Ifg}
&&I(f,g)=\int_{\Om}\left\{\frac1{2\alpha}|\nabla f|^2+\frac1{2\beta}|\nabla
g|^2+(N-1)\e^{\frac12 (u_0+v_0)+f}\right.\nn\\
&&\left.\quad
+\frac12\left(\e^{u_0+f+g}+\e^{v_0+f-g}\right)+\left(\frac{2\pi
}{\alpha|\Om|}[m+n]-\frac\gamma2\right)f+\frac{2\pi
}{\beta|\Om|}[m-n]g\right\}\,\dd x.\quad\quad\quad
\eea

On the other hand, integrating the equations (\ref{7f}) and (\ref{7g}), we
obtain the natural constraints
\bea
\int_\Om\left\{(N-1)\e^{\frac12 (u_0+v_0)+f}+\e^{u_0+f+g}\right\}\,\dd
x&=&\frac\gamma2|\Om|-\frac{2\pi}\alpha (m+n)-\frac{2\pi}\beta (m-n)\nn\\
&\equiv& \eta_1>0,\label{cs1}\\
\int_\Om\left\{(N-1)\e^{\frac12 (u_0+v_0)+f}+\e^{v_0+f-g}\right\}\,\dd
x&=&\frac\gamma2|\Om|-\frac{2\pi}\alpha (m+n)+\frac{2\pi}\beta (m-n)\nn\\
&\equiv& \eta_2>0.\label{cs2}
\eea
In terms of the quantities $\eta_1$ and $\eta_2$ in (\ref{cs1}) and
(\ref{cs2}), we may rewrite (\ref{Ifg}) as
\bea
I(f,g)&=&\int_{\Om}\left\{\frac1{2\alpha}|\nabla
\dot{f}|^2+\frac1{2\beta}|\nabla \dot{g}|^2+(N-1)\e^{\frac12
(u_0+v_0)+f}\right\}\,\dd x\nn\\
&&
+\frac12\left(\int_\Om\left\{\e^{u_0+\ud{U}+\dot{U}}+\e^{v_0+\ud{V}+\dot{V}}\right\}\,\dd
x-\eta_1\ud{U}-\eta_2\ud{V}\right).
\eea
In view of the analysis presented earlier (e.g., \S 3.1 and
(\ref{Iv})--(\ref{2.21}) in particular), we see that the existence of a
critical point of the functional (\ref{Ifg}) follows as a consequence
of the condition $\eta_1,\eta_2>0$.

Finally, adding $\eta_1$ and $\eta_2$, we have
\be
\eta\equiv\gamma|\Om|-\frac{4\pi}\alpha(m+n)>0.
\ee
 Thus, in view of $\eta_1,\eta_2>0$ again, we have
\be
-\eta<\frac{4\pi}\beta(m-n)<\eta,
\ee
which leads to the single condition (\ref{7.5}). On the other hand, it is
obvious that (\ref{7.5}) also implies $\eta_1,\eta_2>0$. Thus the proof of
the theorem follows.

\subsection{Vortices in a softly broken SQCD model}

In this subsection, we construct multiple vortices in the SQCD model of
Marshakov and Yung \cite{MY} in which the confinement is achieved through
Abelian fluxes generated in the Cartan subalgebra sector of $SU(3)$ so that,
in terms of the Gell-Mann matrices $\hat{\lm}_3$ and $\hat{\lm}_8$,
the gauge field $A_\mu$ assumes the form
\be
A_\mu=A_\mu^{(3)}\hat{\lm}_3+A_\mu^{(8)}\hat{\lm}_8,
\ee
where $A_\mu^{(3)}$ and $A_\mu^{(8)}$ are two real-valued vector fields. As
in \cite{MY}, we use u and d to denote the up and down colors of quarks,
which are represented by
a pair of complex-valued Higgs scalar fields, say $\phi^{(\mbox{u})}$ and
$\phi^{(\mbox{d})}$, respectively. The u- and d-fluxes will be the Cartan
subalgebra valued
which are induced from the
gauge fields
\be
A_\mu^{(\mbox{u})}=\frac{\sqrt{3}}2 A_\mu^{(3)}+\frac12 A_\mu^{(8)},\quad
A_\mu^{(\mbox{d})}= -\frac{\sqrt{3}}2 A_\mu^{(3)}+\frac12 A_\mu^{(8)}.
\ee

With the above notation, the effective action functional of the SQCD model
of Marshakov and Yung \cite{MY} reads
\bea
S
&=&\int\left\{\frac1{4g^2}F_{\mu\nu}^{(3)}F^{(3)\mu\nu}+\frac1{4g^2}F_{\mu\nu}^{(8)}F^{(8)\mu\nu}\right.\nn\\
&&+\left(\overline{\nabla_\mu^{(\mbox{u})}\phi^{(\mbox{u})}}\right)
\nabla^{(\mbox{u})\mu}\phi^{(\mbox{u})}
+\left(\overline{\nabla_\mu^{(\mbox{d})}\phi^{(\mbox{d})}}\right)
\nabla^{(\mbox{d})\mu}\phi^{(\mbox{d})}\nn\\
&&+\left.\frac{g^2}8\left(|\phi^{(\mbox{u})}|^2-|\phi^{(\mbox{d})}|^2-\xi[1-\omega]\right)^2+\frac{g^2}{24}\left(|\phi^{(\mbox{u})}|^2+|\phi^{(\mbox{d})}|^2-\xi[1+\omega]\right)^2\right\}\,\dd
x,\nn\\
\eea
where $g,\xi,\omega>0$ are all physical constants, and
\be
\nabla_\mu^{\mbox{(u,d)}}=\pa_\mu -\frac{\ii}{\sqrt{3}}
A^{\mbox{(u,d)}}_\mu,\quad \mu=0,1,2,3,
\ee
are gauge-covariant derivatives to be operated upon $\phi^{\mbox{(u,d)}}$,
accordingly. Concentrating on static case for which the field configurations
are uniform
in a spatial direction, say $x^3$, we see that the method of Bogomol'nyi
\cite{Bo} may be used to show that the equations of motion may be reduced
into the following
BPS \cite{Bo,PS} system \cite{MY}
\bea
\nabla^{\mbox{(u,d)}}_1\phi^{\mbox{(u,d)}}+\ii
\nabla^{\mbox{(u,d)}}_2\phi^{\mbox{(u,d)}}&=&0,\label{8a}\\
F^{(3)}_{12}+\frac{g^2}2\left(|\phi^{(\mbox{u})}|^2-|\phi^{(\mbox{d})}|^2-\xi[1-\omega]\right)&=&0,\label{8b}\\
F^{(8)}_{12}+\frac{g^2}{2\sqrt{3}}\left(|\phi^{(\mbox{u})}|^2+|\phi^{(\mbox{d})}|^2-\xi[1+\omega]\right)&=&0,\label{8c}
\eea
subject to the boundary condition
\be
|\phi^{\mbox{(u)}}|^2\to\xi,\quad |\phi^{\mbox{(d)}}|^2\to\xi\omega,
\ee
so that the (minimum) vortex-line energy or tension may be calculated via
the flux formula
\be \label{tension}
T=\frac\xi{\sqrt{3}}\int_{\bfR^2} \left(F_{12}^{\mbox{(u)}}+\omega
F_{12}^{\mbox{(d)}}\right)\,\dd x.
\ee

With the notation
\be
\phi=\phi^{\mbox{(u)}},\quad \psi=\phi^{\mbox{(d)}},\quad a_\mu=\frac12
A_\mu^{(3)}+\frac{\sqrt{3}}2A_\mu^{(8)},\quad b_\mu=A_\mu^{(\mbox{d})},
\ee
and the relations
\be
A^{\mbox{(u)}}_\mu=\frac{\sqrt{3}}2a_\mu-\frac12 b_\mu,\quad
\nabla^{\mbox{(u)}}_\mu
=\pa_\mu-\frac{\ii}2\left(a_\mu-\frac1{\sqrt{3}}b_\mu\right),\quad
\nabla_\mu^{\mbox{(d)}}=\pa_\mu-\frac{\ii}{\sqrt{3}} b_\mu,
\ee
we may rewrite the equations (\ref{8a})--(\ref{8c}) as
\bea
(\pa_1+\ii\pa_2)\phi&=&\frac{\ii}2\left([a_1+\ii
a_2]-\frac1{\sqrt{3}}[b_1+\ii b_2]\right)\phi,\label{8d}\\
(\pa_1+\ii\pa_2)\psi&=&\frac{\ii}{\sqrt{3}}(b_1+\ii b_2)\psi,\\
a_{12}&=&\frac{g^2}2(\xi-|\phi|^2),\\
b_{12}&=&\frac{g^2}{\sqrt{3}}\left(\xi\left[\omega-\frac12\right]+\frac12|\phi|^2-|\psi|^2\right).\label{8e}
\eea

Assume the sets of zeros of $\phi$ and $\psi$ are as prescribed in
(\ref{Zpq}). Then, as before, the substitution, $u=\ln|\phi|^2$ and
$v=\ln|\psi|^2$, allows us to transform
the equations (\ref{8d})--(\ref{8e}) into
\bea
\Delta u+\frac{1}2\Delta
v&=&\frac{g^2}2(\e^u-\xi)+4\pi\sum_{s=1}^m\delta_{p_s}(x)+2\pi
\sum_{s=1}^n\delta_{q_s}(x),\label{8.40}\\
\Delta
v&=&\frac{g^2}{3}\left(-\e^u+2\e^v-\xi[2\omega-1]\right)+4\pi\sum_{s=1}^n\delta_{q_s}(x).\label{8.41}
\eea

We consider a doubly periodic domain $\Om$ first. Let $u_0$ and $v_0$ be
given in (\ref{uv00}). Then $u=u_0+U$ and $v=v_0+V$ recast
(\ref{8.40})--(\ref{8.41}) into
\bea
\Delta U+\frac{1}2\Delta V&=&\frac{g^2}2(\e^{u_0+U}-\xi)+\frac{4\pi
m}{|\Om|}+\frac{2\pi n}{|\Om|},\label{8.42}\\
\Delta
V&=&\frac{g^2}{3}\left(-\e^{u_0+U}+2\e^{v_0+V}-\xi[2\omega-1]\right)+\frac{4\pi
n}{|\Om|}.\label{8.43}
\eea
Set $W=U+\frac{1}2 V$. We arrive at
\bea
\Delta V&=&\frac{g^2}{3}\left(2\e^{v_0+V}-\e^{u_0-\frac{1}2 V+W}-\xi[2\omega
-1]\right)+\frac{4\pi n}{|\Om|},\label{8eq1}\\
\Delta W&=&\frac{g^2}2\left(\e^{u_0-\frac{1}2
V+W}-\xi\right)+\frac{2\pi}{|\Om|}(2m+ n),\label{8eq2}
\eea
which are the Euler--Lagrange equations of the functional
\bea\label{IVW}
I(V,W)&=&\int_\Om\left\{\frac3{4g^2}|\nabla V|^2+\frac1{g^2}|\nabla
W|^2+\e^{u_0-\frac{1}2 V+W}+\e^{v_0+V}\right.\nn\\
&& \left.  -\left(\xi\left[\omega-\frac12\right]-\frac{6\pi
n}{g^2|\Om|}\right)V-\left(\xi-\frac{4\pi}{g^2|\Om|}[2m+n]\right)W
\right\}\,\dd x.\nn\\
\eea

To proceed further, we integrate (\ref{8eq1}) and (\ref{8eq2}) to obtain the
constraints
\bea
\int_\Om \e^{u_0-\frac{1}2 V+W}\,\dd
x&=&\xi|\Om|-\frac{4\pi}{g^2}(2m+n)\nn\\
&\equiv&\eta_1>0,\label{8ca}\\
\int_\Om \e^{v_0+V}\,\dd x&=& \xi\omega|\Om|-\frac{4\pi}{g^2}(m+2n)\nn\\
&\equiv&\eta_2>0.\label{8cb}
\eea

Therefore, we have
\bea \label{8I}
&&I(V,W)-\frac1{g^2}\int_\Om\left\{\frac34|\nabla\dot{V}|^2+|\nabla\dot{W}|^2\right\}\,\dd
x\nn\\
&&=\left(\int_\Om \e^{u_0+\ud{U}+\dot{U}}\,\dd x
-\eta_1\ud{U}\right)+\left(\int_\Om \e^{v_0+\ud{V}+\dot{V}}\,\dd
x-\eta_2\ud{V}\right).
\eea
In view of (\ref{8ca})--(\ref{8I}), we see that the existence of a critical
point, in fact, a global minimizer, of the functional (\ref{IVW}) in
$W^{1,2}(\Om)$ follows as before.

We now consider (\ref{8.40}) and (\ref{8.41}) over the full plane. In
(\ref{Zpq}), use the notation
\be
p_{s}=p_{1,s},\quad s=1,\cdots,m\equiv n_1;\quad q_s=p_{2,s},\quad
s=1,\cdots, n\equiv n_2.
\ee
Let $u_\ell^0$ and $h_\ell$ be defined as in (\ref{u0}) and (\ref{h}),
respectively, $\ell=1,2$. Introduce the translations
\be
u=\ln\xi+u^0_1+v_1,\quad v=\ln(\xi\omega)+u^0_2+v_2,
\ee
and the refined parameters
\be
\alpha=\quad \frac12g^2\xi,\quad \beta=\frac13{g^2}\xi,\quad
\gamma=\frac23{g^2}\xi\omega.
\ee
We do so since the ranges of these parameters will not be important for our
existence theory over $\bfR^2$. The equations (\ref{8.40}) and (\ref{8.41})
now become
\bea
\Delta v_1 +\frac12\Delta
v_2&=&\alpha\left(\e^{u_1^0+v_1}-1\right)+h_1+\frac{1}2 h_2,\\
\Delta
v_2&=&-\beta\left(\e^{u_1^0+v_1}-1\right)+\gamma\left(\e^{u_2^0+v_2}-1\right)+h_2.
\eea
Set $v_1+\frac12 v_2=w_1, v_2=w_2$. Then we have the following modified
system of equations
\bea
\Delta w_1 &=&\alpha\left(\e^{u_1^0+w_1-\frac12 w_2}-1\right)+h_1+\frac{1}2
h_2,\\
\Delta w_2&=&-\beta\left(\e^{u_1^0+w_1-\frac12
w_2}-1\right)+\gamma\left(\e^{u_2^0+w_2}-1\right)+h_2,
\eea
which are the Euler--Lagrange equations of the functional
\bea \label{Iww}
I(w_1,w_2)&=&\int_{\bfR^2}\left\{\frac1{2\alpha}|\nabla
w_1|^2+\frac{1}{4\beta}|\nabla w_2|^2+\e^{u^0_1 +w_1-\frac12
w_2}-\e^{u^0_1}-\left[w_1-\frac12 w_2\right]\right.\nn\\
&&\left.+\frac{\gamma}{2\beta}\left(\e^{u^0_2+w_2}-\e^{u^0_2}-w_2\right)+\frac1\alpha\left(h_1+\frac{1}2
h_2\right)w_1+\frac 1{2\beta} h_2 w_2\right\}\,\dd x.\nn\\
\eea
It is clear that this functional is $C^1$ and strictly convex over
$W^{1,2}(\bfR^2)$. Besides, we have
\bea
&&(DI(w_1,w_2))(w_1,w_2)=\int_{\bfR^2}\left\{\frac1\alpha|\nabla
w_1|^2+\frac 1{2\beta}|\nabla w_2|^2\right\}\,\dd x\nn\\
&&+\int_{\bfR^2}\left\{\left(\e^{u_1^0+w_1-\frac12
w_2}-1\right)\left(w_1-\frac12
w_2\right)+\frac{\gamma}{2\beta}\left(\e^{u^0_2+w_2}-1\right)w_2\right.\nn\\
&&\left. +\frac1\alpha\left(h_1+\frac{1}2 h_2\right)w_1+\frac 1{2\beta} h_2
w_2\right\}\,\dd x.
\eea
Thus, we can show that there are constants $C_1,C_2>0$ such that
\be
(DI(w_1,w_2))(w_1,w_2)\geq
C_1\left(\|w_1\|_{W^{1,2}(\bfR^2)}+\|w_2\|_{W^{1,2}(\bfR^2)}\right)-C_2,\quad\forall
w_1, w_2.
\ee
Consequently, the existence and uniqueness of a critical point of the
functional (\ref{Iww}) over $W^{1,2}(\bfR^2)$ is established.

In summary, we may state

\begin{theorem}
Consider the BPS system of SQCD vortex equations (\ref{8d})--(\ref{8e}) for
$(\phi,\psi,a_j,b_j)$ with the prescribed sets
of zeros given in (\ref{Zpq}) so that $\phi,\psi$ have $m,n$ arbitrarily
distributed zeros, respectively.

(i) For this problem over a doubly periodic domain $\Om$, a solution exists
if and only if the  condition
\be \label{8cc}
\max\left\{\frac1\omega(m+2 n),2m+n\right\}<\frac{g^2\xi |\Om|}{4\pi}
\ee
holds. Moreover, whenever a solution exists, it is unique.

(ii) For this problem over the full plane $\bfR^2$ subject to the
finite-energy boundary condition
\be
|\phi|^2\to\xi,\quad |\psi|^2\to\xi\omega,\quad\mbox{as }|x|\to\infty,
\ee
there exists a unique solution up to gauge transformations so that the
boundary behavior stated above is realized exponentially rapidly.

In either case, the excited total vortex fluxes are quantized quantities
given explicitly by the formulas
\be
\Phi_a=\int a_{12}\,{\rm \dd} x=2\pi(2 m + n),\quad \Phi_b=\int
b_{12}\,{\rm\dd} x=2\sqrt{3}\pi n,
\ee
respectively, and the solutions may be obtained by  methods of direct
minimization.
\end{theorem}

We note that the condition (\ref{8cc}) is simply a suppressed restatement of
the two simultaneous conditions (\ref{8ca}) and (\ref{8cb}).

We also note that, applying the relations between the gauge fields $a_\mu,
b_\mu$ and $A_\mu^{(3)}, A_\mu^{(8)}$, we easily obtain
the Abelian (or the Cartan subalgebra valued) fluxes
\be
\Phi^{(3)}=\int F^{(3)}_{12}\,\dd x=2\pi(m-n),\quad \Phi^{(8)}=\int
F_{12}^{(8)}\,\dd x=2\sqrt{3}\pi(m+n).
\ee
Moreover, the u- and d- fluxes may be expressed by the formulas
\be
\Phi^{(\mbox{u})}=\int F_{12}^{(\mbox{u})}=2\sqrt{3}\pi m,\quad
\Phi^{(\mbox{d})}=\int F_{12}^{(\mbox{d})}=2\sqrt{3}\pi n,
\ee
which depend on the winding numbers of the u- and d-Higgs fields
$\phi^{(\mbox{u})}$ and $\phi^{(\mbox{d})}$, respectively, and give rise to
the tension or energy of the vortex-lines
\be
T=2\pi (m+\omega n)\xi,
\ee
according to (\ref{tension}).

\medskip

We also note that our direct method works easily for the classical BPS
Abelian Higgs vortex equations \cite{JT} defined over doubly periodic
domains \cite{WY} or formulated
over compact Riemann surfaces \cite{Nog1,Nog2}.

\end{document}